\documentclass [prb,reprint,notitlepage,groupedaddress,floatfix] {revtex4-2}
\usepackage{amsmath}
\usepackage{graphicx}
\usepackage{lmodern}
\usepackage{amssymb}
\usepackage[caption=false]{subfig}
\usepackage[colorlinks=true, citecolor=green, urlcolor=blue, linkcolor=red]{hyperref}

\renewcommand{\(}{\left (}
\renewcommand{\)} {\right )}
\renewcommand{\]} {\right ]}
\renewcommand{\[} {\left [}

\begin{document}

\title{{Nodal higher-order topological superconductivity from a $\mathcal{C}_4$-symmetric Dirac semimetal}}
\author{Zhenfei Wu}
\author{Yuxuan Wang}
\affiliation{Department of Physics, University of Florida, Gainesville, Florida, 32611, USA}
\date{\today}

\begin{abstract}
We analyze the topological properties of the possible superconducting states emerging from a Cd$_3$As$_2$-like, $\mathcal{C}_4$-symmetric Dirac semimetal, with two fourfold-degenerate Dirac points separated in the $k_z$ direction. Unlike the simplest Weyl semimetal for which all pairing orders are topologically obstructed and nodal, we show that the topological obstruction for pairing in Dirac semimetals crucially only exists for certain pairing symmetries. In particular, we focus on odd-parity $B_{1u}$ and $B_{2u}$ pairing states, both of which can be induced by Ising ferromagnetic fluctuations. The $B_{1u}$ and $B_{2u}$ pairing states inherit the topological obstruction from the normal state, which dictates that these states necessarily host four Bogoliubov-de Gennes (BdG) Dirac point nodes protected by a $\mathbb{Z}_2$ monopole charge. By a Wannier state analysis,  we show that the topological obstruction in the superconducting states is of higher-order nature. As a result, in a rod geometry with gapped surfaces, arcs of higher-order Majorana zero modes exist in certain $k_z$ regions of the hinges between the BdG Dirac points. Unlike Fermi arcs in Weyl semimetals, the higher-order Majorana arcs are stable against self-annihilation due to an additional $\mathbb{Z}$-valued monopole charge of the BdG Dirac points protected by $\mathcal{C}_4$ symmetry. We find that the same $\mathbb{Z}$-valued charge is also carried by $B_{1g}$ and $B_{2g}$ channels, where the BdG spectrum hosts bulk ``nodal cages," i.e., cages formed by nodal lines, that are stable against symmetry-preserving perturbations.
\end{abstract}

\maketitle

%\tableofcontents

\section{Introduction}
Topological semimetals are gapless states that exhibit topologically stable band crossings in the Brillouin zone (BZ), including Weyl \cite{Weyl1,Weyl2,Weyl4,Weyl6,Weyl3,Weyl5,Weyl7,Weyl8,Weyl9,Weyl10,Weyl11} and Dirac semimetals \cite{DSM1,DSM2,DSM3,DSM4,DSM5,DSM6,DSM7}, nodal line semimetals, \cite{NLSM1,NLSM2,NLSM3,NLSM4,NLSM7,Nodalcage,NLSM5,NLSM6,BJY_spin_polarized} and multifold fermions \cite{multi1,multi2,multi3}. {Owing to the nontrivial topology, topological semimetals display interesting spectral and transport properties.} For example, in Weyl semimetals there exist gapless surface states that are restricted in certain ranges in momentum space, known as Fermi arcs, which leads to interesting features seen in ARPES and quantum oscillation measurements \cite{Weyl1,Weyl2,Weyl4,Weyl6}.

Moreover, topological semimetals are fertile playgrounds for the interplay between correlation effects and band topology. In particular, interacting topological semimetals can host novel superconducting instabilities upon doping, whose topological properties descend from the normal state \cite{MonopoleWeyl,SatoWSC}. For example, in a Weyl semimetal with two Weyl points, it has been shown that \emph{all} possible pairing orders are necessarily topologically obstructed, i.e., hosting point nodes in the Bogoliubov-de Gennes (BdG) spectrum \cite{MonopoleWeyl}. The topological obstruction comes from a Chern number for surfaces enclosing the Weyl points, which is inherited by the superconducting state. As a consequence, the surface Fermi arcs in the normal states get reconnected in the superconducting states and terminate at the BdG point nodes instead \cite{SatoWSC}.

Recently, the notion of band topology has been extended to higher-order ones, with examples ranging from higher-order topological insulators \cite{Hughes_corner,BBH_Science,Hughes_corner_PRB,frac_corner_SOC,SongHOTI,HOTI_Green_function}, higher-order topological superconductors \cite{BJY_spin_polarized,JahinTiwariWang,Wang_Hughes,TiwariJahinWang,HOTSC_mixed,HO_reflection,HO_inversion,HOTSC_Hubbard,Hightemp_majorana,HOTSC_sr2ruo4,Majorana_HOTI,Majorana_2DTI_SC,HOTSC_pi,hingeMajorana_FeSC,HOTSC_oddP,HOTSC_lattice,HOTSC_nodalSC,HOTSC_2D,HOTSC_wte2,HOTSC_quadrupolarDSM,octupolar,HOTSC_C2,Vortexline_FeSC,HOTSCin3D,HOTSC_weakTI,HOTSC_spin_polarization,qin_fang_zhang_hu,qin_zhang_wang_fang_zhang_hu}, and higher-order topological semimetals \cite{quadrupolar_SM,HOWSM,Zhou_Ye,NH_HOWSM,NH_HODSM,sub_topology}. In general, $n$th-order higher-order topological phases are protected by spatial symmetries and host gapless or degenerate states on its $(d - n)$-dimensional boundary ($1 < n \leq d$) while its bulk and surface spectra are gapped. As examples, interesting higher-order topological superconductivity has been shown to emerge from nodal line semimetals \cite{BJY_spin_polarized}, time-reversal-invariant Weyl semimetals \cite{JahinTiwariWang}, and Dirac semimetals \cite{Wang_Hughes,TiwariJahinWang}.

Different from Weyl semimetals \cite{Weyl1,Weyl2,Weyl4,Weyl6,Weyl3,Weyl5}, the topological stability of Dirac semimetals requires additional symmetry. In Dirac semimetals such as Cd$_3$As$_2$ and Au$_2$Pb, the Dirac points are protected by a $\mathbb{Z}_2$ topological invariant in the presence of time-reversal, inversion, and fourfold-rotational $\mathcal{C}_4$ symmetries \cite{SatoSC,DSM_BJY_charge,DSM_BJY_classify}. In terms of bulk-boundary correspondence, Dirac semimetals do not host stable Fermi arcs on the surfaces \cite{DSMnoFA1,DSMnoFA2,DSMnoFA3,SatoSC}. Instead, recent studies have shown that certain Dirac semimetals display higher-order Fermi arcs (HOFAs) \cite{NC_HOFA,Fang_classifyHOFA,Fang_fillinganomaly}, which are one-dimensional (1D) degenerate dispersive states localized on the lower-dimensional hinges when two open boundaries intersect \cite{quadrupolar_SM,HOWSM,SongHOTI,Zhou_Ye}. The HOFAs  terminate at the projections of bulk Dirac points onto 1D hinges, and are protected by the same symmetries protecting the bulk Dirac points.

Superconductivity has been reported in Dirac materials such as Cd$_3$As$_2$ \cite{SC_cd3as2_exp1,SC_cd3as2_exp2,SC_cd3as2_exp3} and Au$_2$Pb \cite{SC_au2pb_exp}, followed by several theoretical investigations of topological superconductivity in doped Dirac semimetals \cite{SatoSC,SC_dopedDSM,monopoleDSM,Sun_Z2}. In this work we investigate possible topologically obstructed nodal superconducting states from a $\mathcal{C}_4$-symmetric Dirac semimetal. If such superconducting states do exist, a natural issue is their indication to the bulk-boundary correspondence. Given that the normal state exhibits nontrivial higher-order topology, such pairing states may display nodal higher-order topological superconductivity. A similar question was addressed in Ref.~\cite{Sun_Z2}. {In that work, the normal state is $\mathbb{Z}_2$ obstructed and the pairing states that inherit the normal state obstruction present nodal points on Fermi surfaces (FSs). The two-dimensional (2D) Hamiltonian is a class DIII topological superconductor due to a 2D composite time-reversal symmetry which satisfies $(\mathcal{TM}_z)^2 = -1$. For bulk-boundary correspondence, {the model studied in Ref.~\cite{Sun_Z2}}  displays surface helical Majorana arcs connecting the projections of bulk gap nodes onto surfaces.} However, Ref.~\cite{Sun_Z2}  considered a mirror reflection symmetry that squares to $+1$, which does not apply to realistic Dirac materials.

Motivated by these questions, we begin with a simple tight-binding model for a doped Dirac semimetal with two Dirac points protected by time-reversal symmetry and the $D_{4h}$ point group \cite{SatoSC}. In the normal state, there are two closed Fermi surfaces around each Dirac point. Such a model can be thought of two copies of Weyl semimetals that are time-reversal partners, with two types of spin-orbit coupling terms $\lambda_1$ and $\lambda_2$ at lowest order in $\bf k$ that couple the two copies. These spin-orbit coupling terms eliminate the surface Fermi arcs and reveal the HOFAs at the hinges.

\begin{table}
    \begin{tabular}{c|c|c}
        \hline
        \hline
        Sample Cut       &  $\vcenter{\hbox{\includegraphics[width=0.4in]{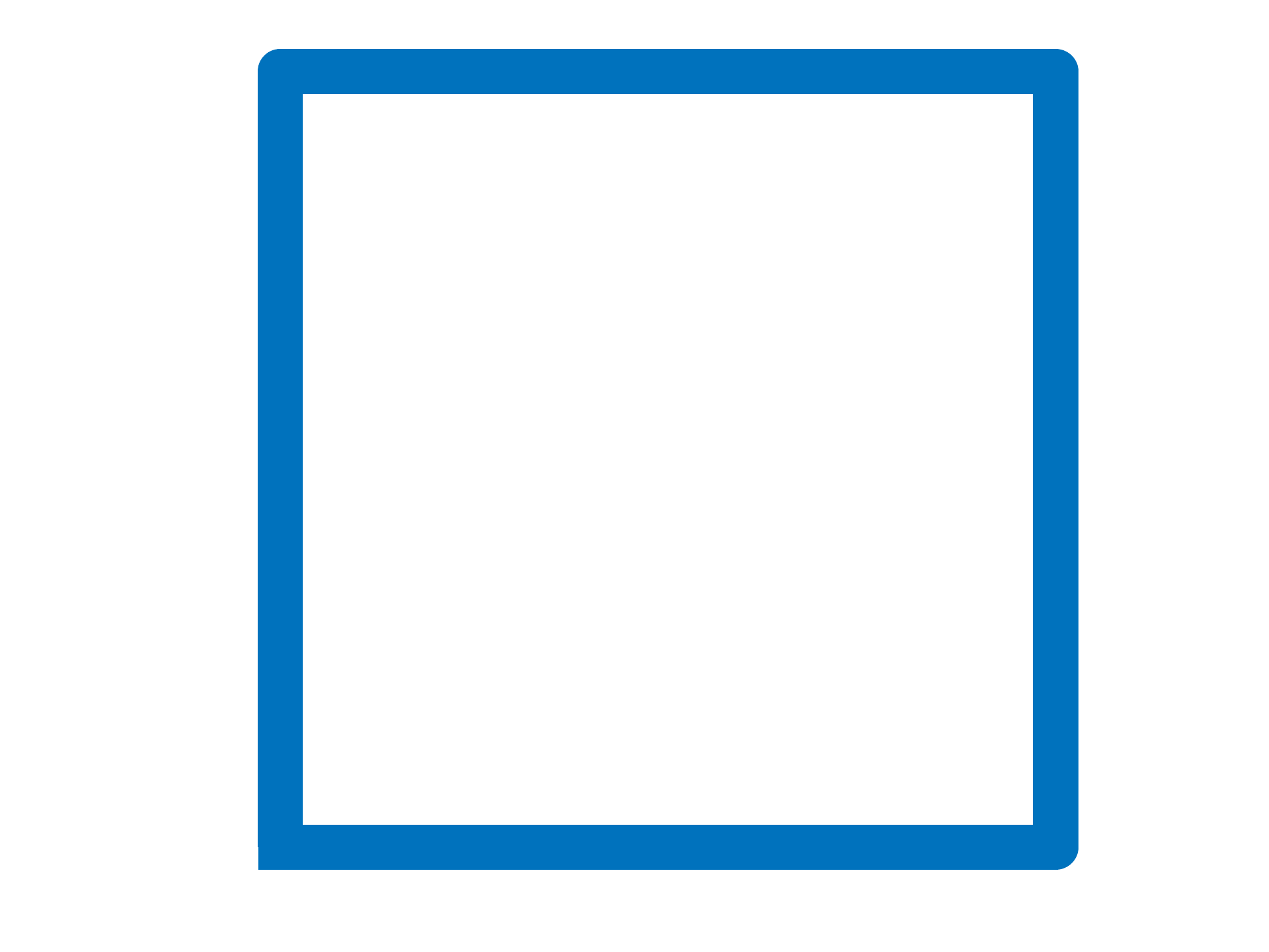}}}$   & $\vcenter{\hbox{\includegraphics[width=0.4in]{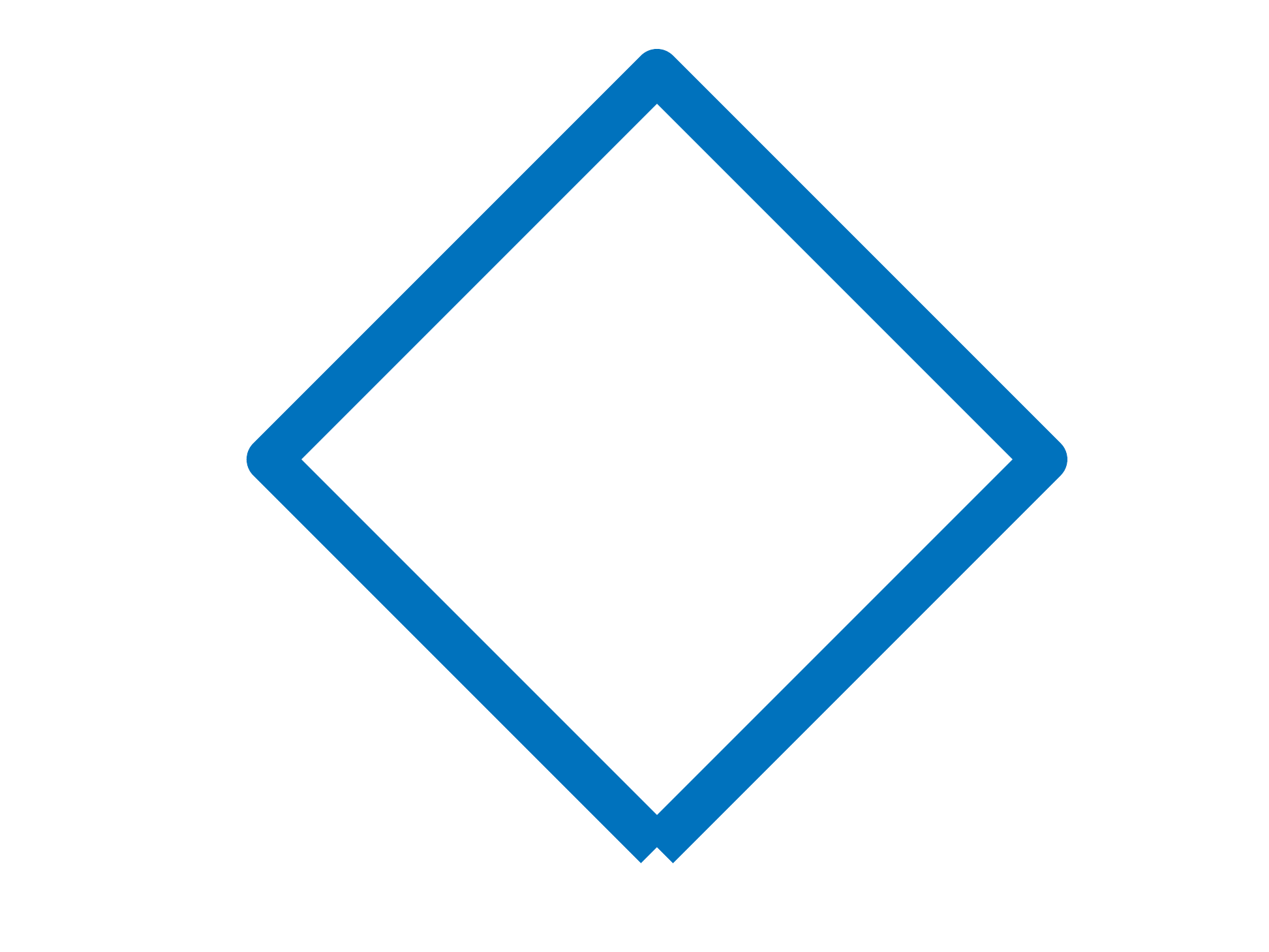}}}$ \\
        \hline
     DSM($k_z \neq 0$)    &  $\vcenter{\hbox{\includegraphics[width=0.4in]{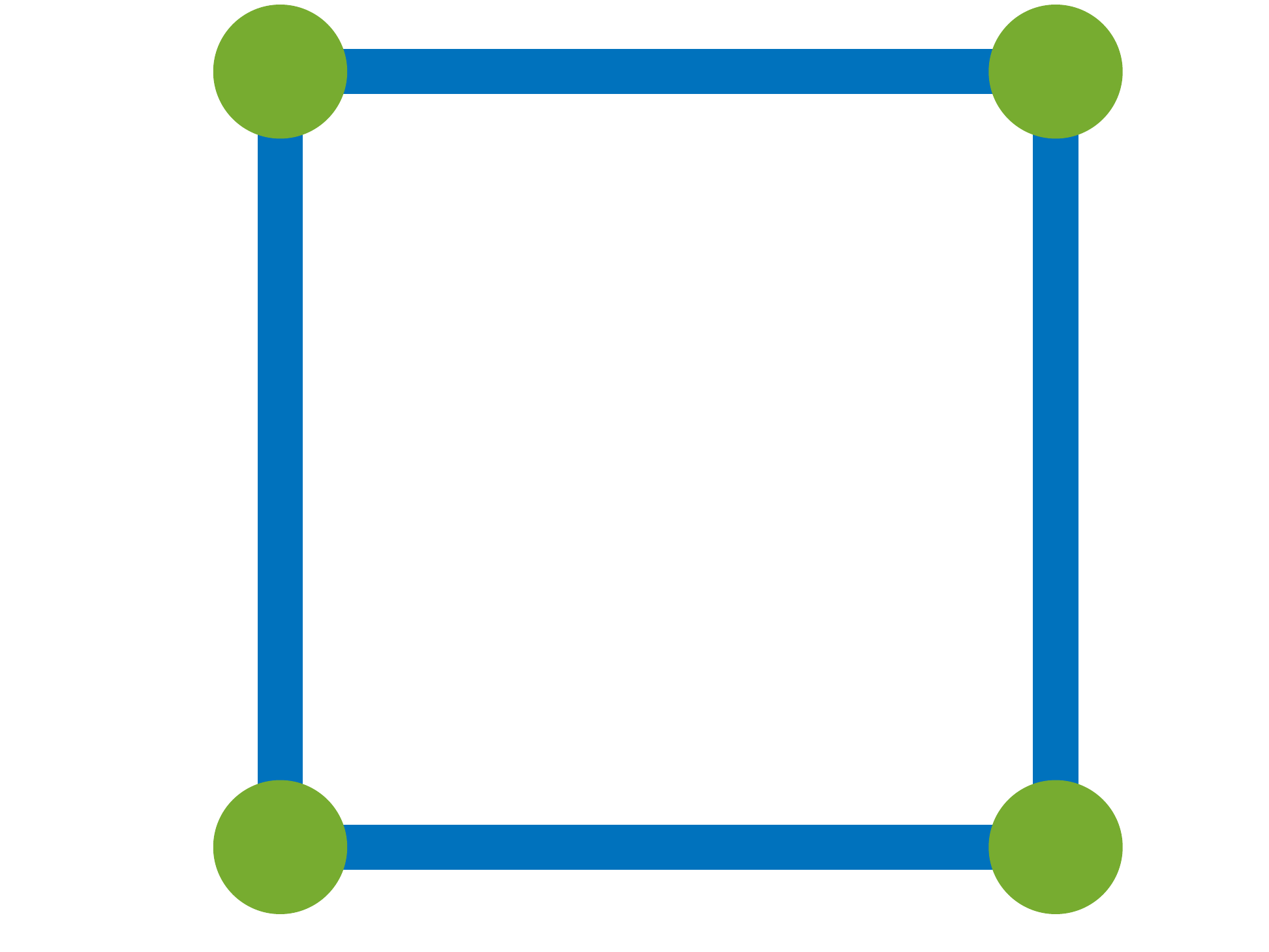}}}$\  HOFAs  &    $\vcenter{\hbox{\includegraphics[width=0.4in]{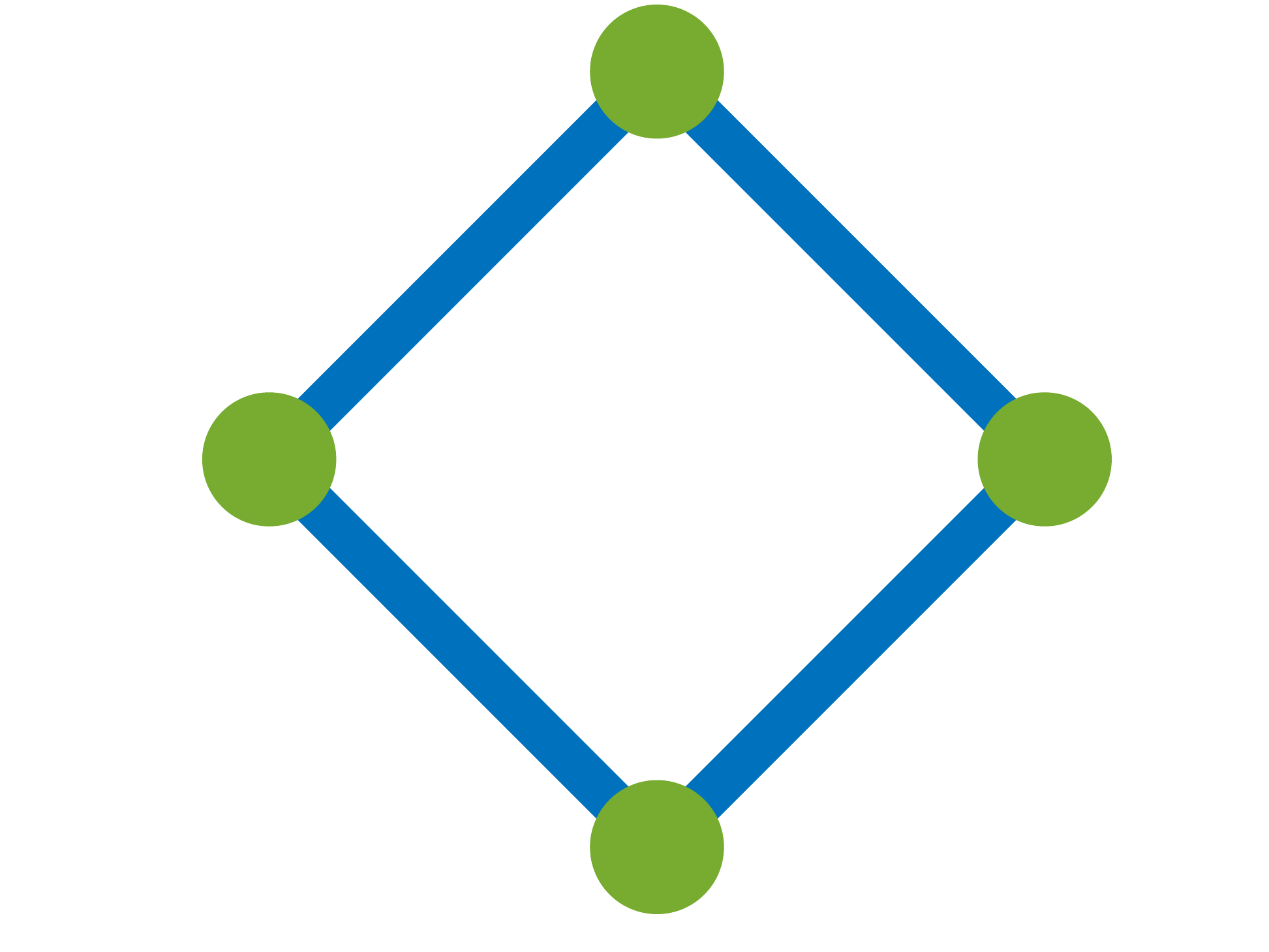}}}$\ HOFAs  \\
        \hline
        $B_{1u}$-DSC  &  $\vcenter{\hbox{\includegraphics[width=0.4in]{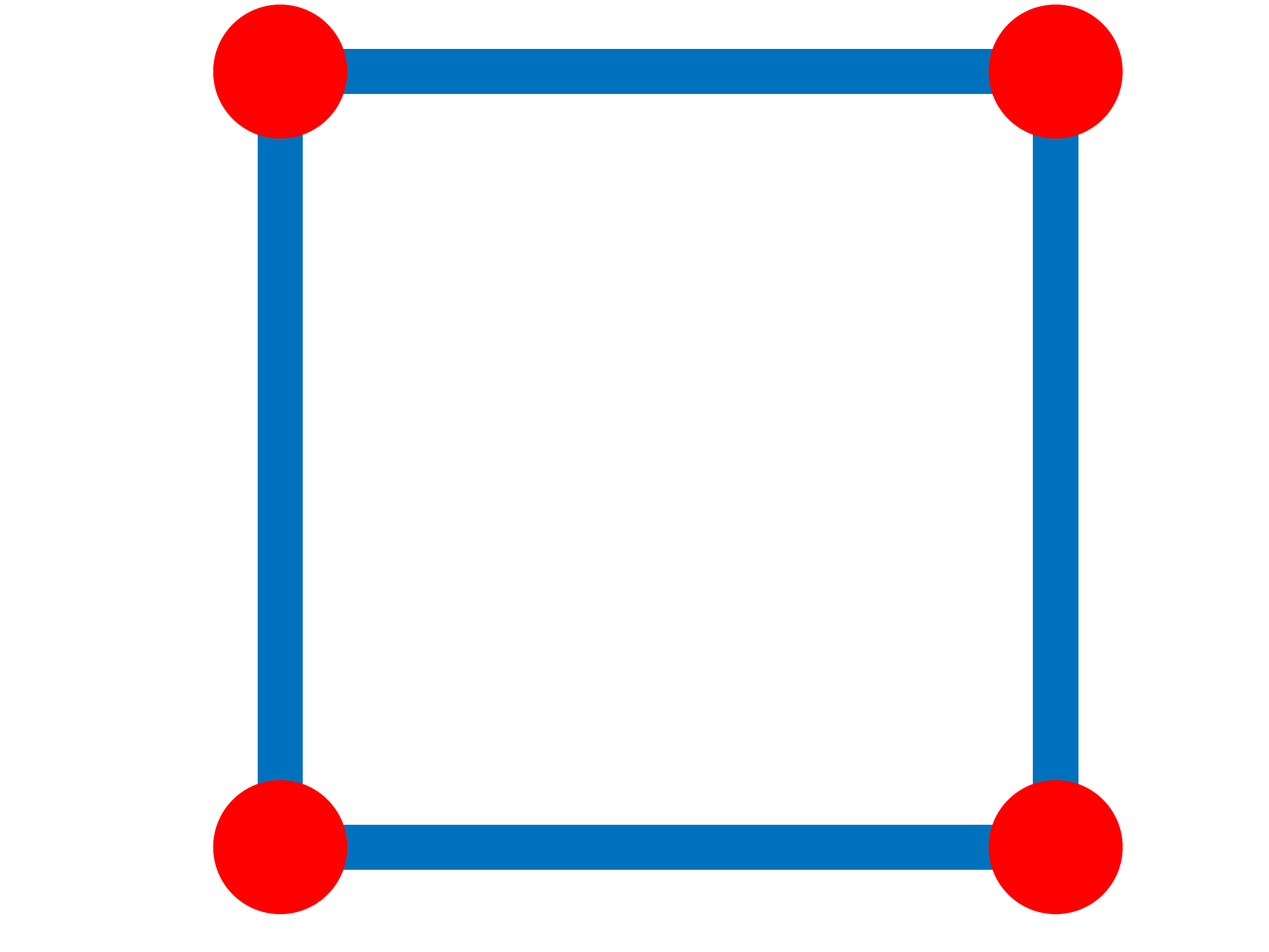}}}$\ HOMAs  &    $\vcenter{\hbox{\includegraphics[width=0.4in]{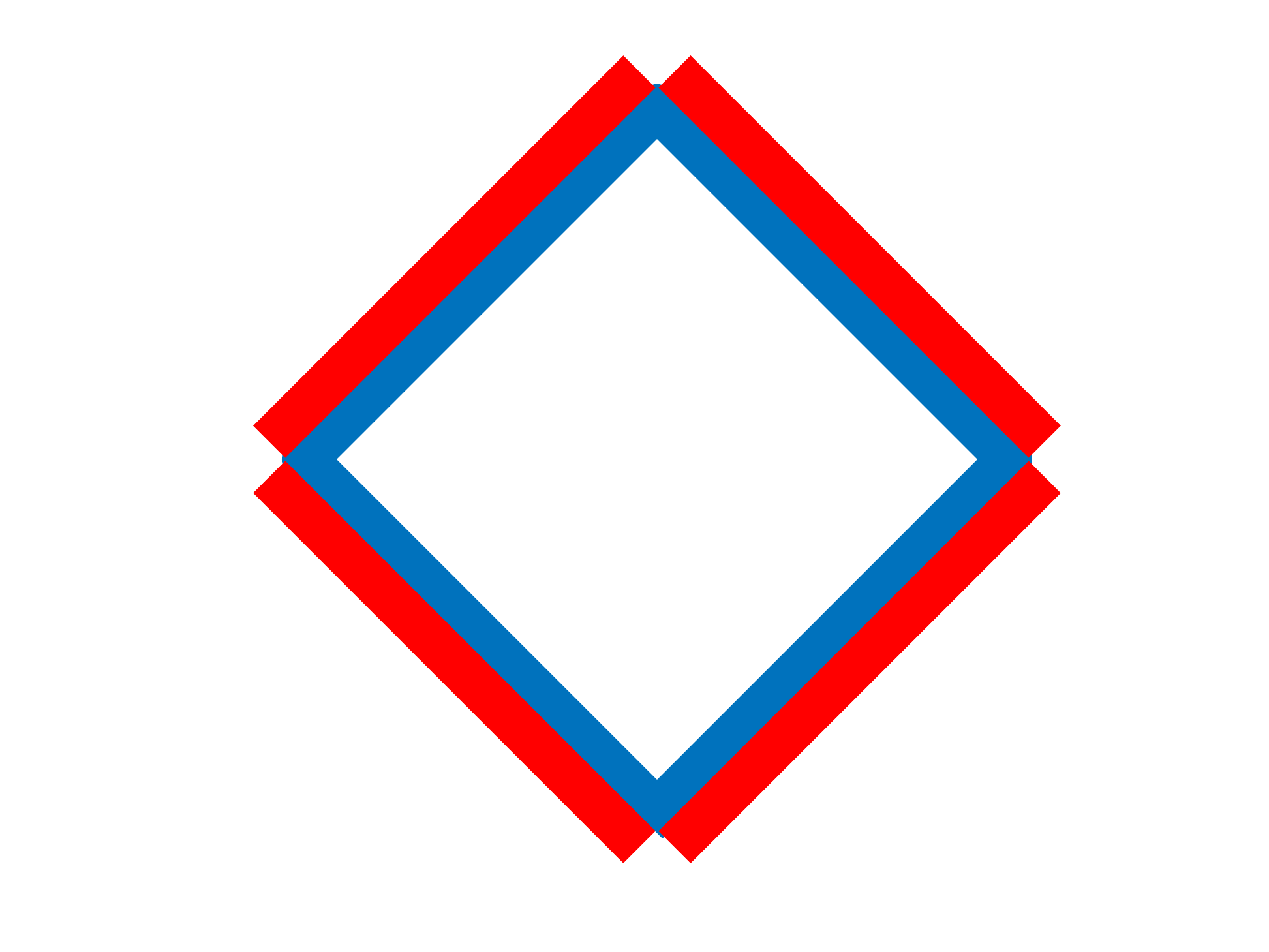}}}$\ gapless  \\
        \hline 
        $B_{2u}$-DSC  &  $\vcenter{\hbox{\includegraphics[width=0.4in]{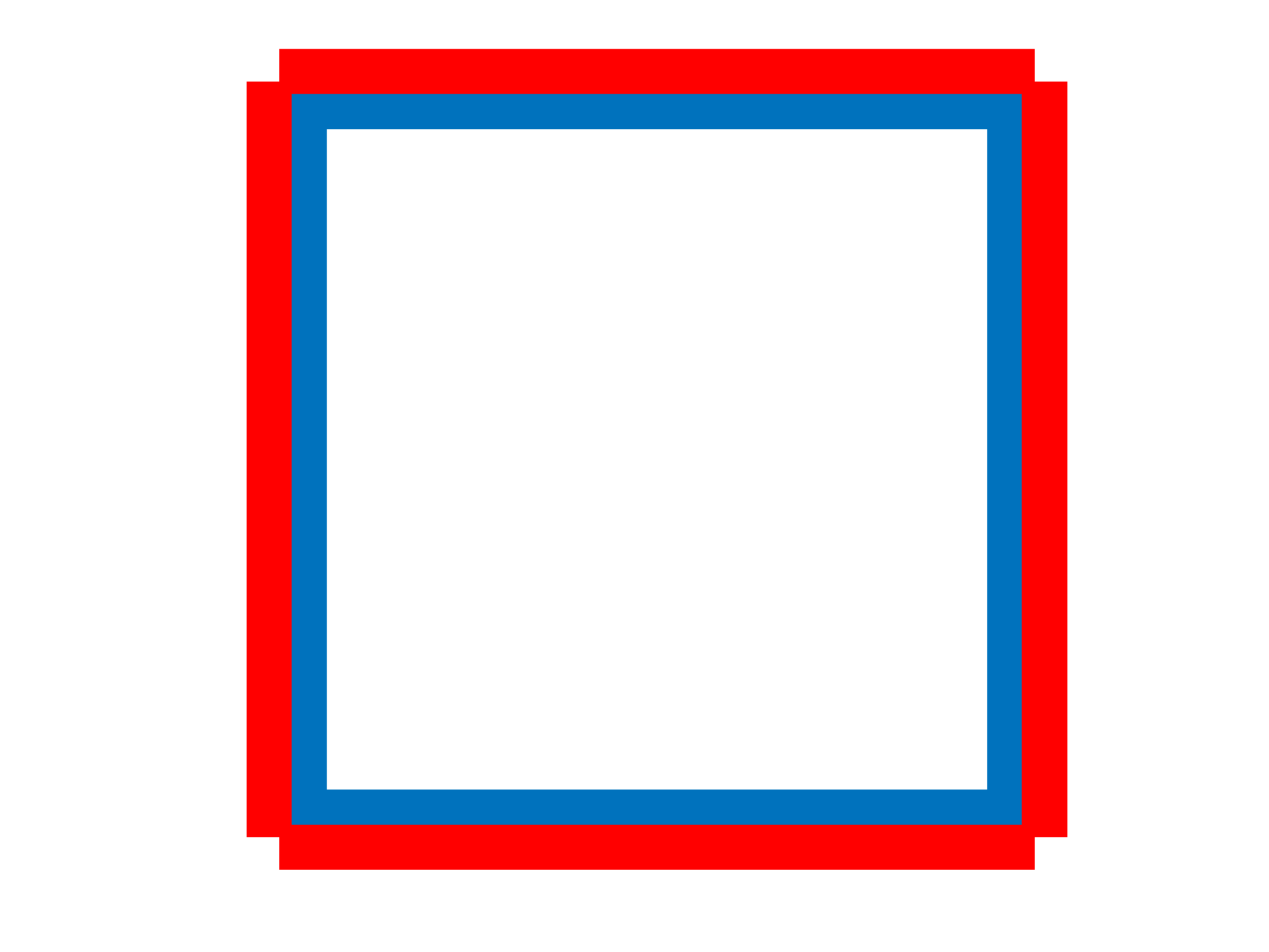}}}$ \ gapless     & $\vcenter{\hbox{\includegraphics[width=0.4in]{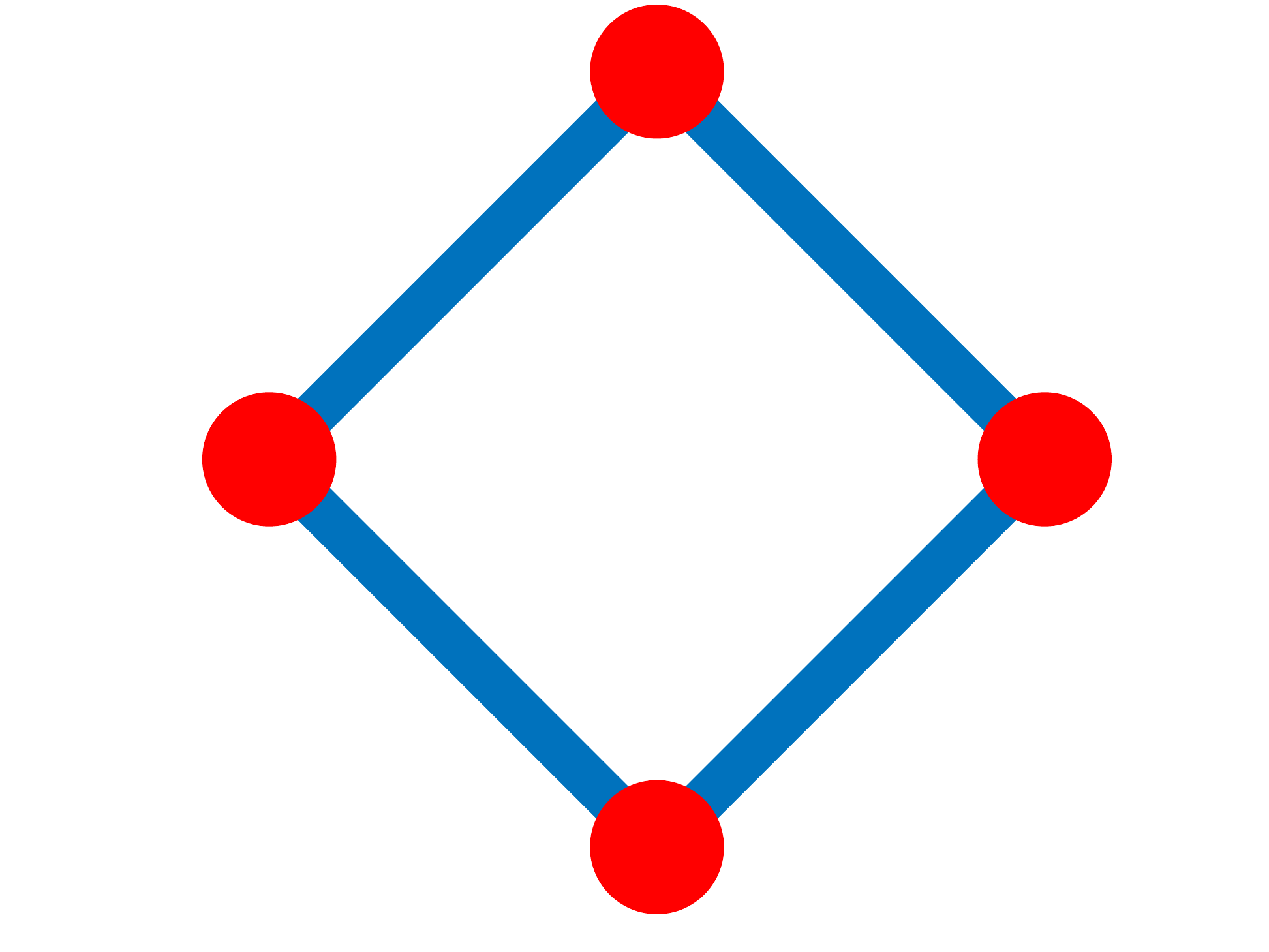}}}$\ HOMAs  \\
        \hline
        \hline
       \end{tabular}
    \caption{Surface states or corner states for a 2D $k_z$ slice in different cases. ``DSM" refers to Dirac semimetal and ``DSC" refers to Dirac superconductor. {The DSM entry excludes the $k_z = 0$ plane, where gapless surface states are present due to a nontrivial mirror Chern number. $B_{1u}$ and $B_{2u}$ are orbital-singlet and spin-triplet pairings given by Eq. \eqref{eq:B1uB2u}.} We use green dots to denote corner states in normal state and red dots to denote corner Majorana zero modes in superconducting state.}
    \label{table:0}
\end{table}

We show that just like the normal state, the topological obstruction of the superconducting state strongly depends on spatial symmetries and, in particular, the irreducible representation thereof~\cite{Sun_Z2}. Out of all possible pairing channels that are irreducible representations of $D_{4h}$, we show that $B_{1u}$ and $B_{2u}$ pairing channels, both of which are odd in parity, inherit the topological obstruction from the normal state. This is in sharp contrast with superconductivity in doped Weyl semimetals with two Weyl points, in which all possible pairing states are necessarily topologically obstructed and nodal \cite{MonopoleWeyl}. We compare the relative strength of these two pairing instabilities. Assuming a scenario with Ising ferromagnetic fluctuations, the effective interaction is attractive for both pairing channels, and whether $B_{1u}$ or $B_{2u}$ is energetically favorable is determined by the relative strength of $\lambda_1$ and $\lambda_2$. Even for a realistic system beyond the simple tight-binding model we consider, we expect the result to hold for small Fermi surfaces.

We further consider the topological nature and the resulting bulk-boundary correspondence of $B_{1u}$ and $B_{2u}$ superconducting states. Crucially, we find that the topological properties are completely determined by the transformation properties of the order parameter under the point group $D_{4h}$. We find that the superconducting order parameters create BdG Dirac points residing on the north and south poles of each of the two disjointed FSs in the BdG Hamiltonian. Higher-order Majorana arcs (HOMAs) localized along the hinges  of a 3D rod geometry (finite in $xy$ but infinite along $z$) exist, originating and terminating at the projections of bulk BdG Dirac points onto the hinges. In the normal state, generally HOFAs are dispersive in the spectrum for a finite chemical potential. By contrast, HOMAs are always pinned at zero energy due to the particle-hole symmetry. We show that the existence of HOMAs can be detected by a $\mathbb{Z}_2$ topological invariant corresponding to the filling anomaly of the 2D subsystems at a given $k_z$ within the relevant range, captured by the positions of the Wannier center of the 2D subsystem. {The model in Ref. \cite{Sun_Z2} has a similar bulk spectra but the higher-order topology is absent since it does not have $\mathcal{C}_4$ symmetry. {In Ref. \cite{Second_order_DSC} the authors proposed a second-order Dirac superconductor with flat-band Majorana arcs localized on hinges. In their model the hinge Majorana arcs terminate at the projections of Dirac nodes from surface states, while our model features bulk Dirac nodes. }}

Interestingly, unlike the normal state, in the pairing state HOMAs only appear for certain vertical hinges, depending on how the finite sample is cut. Preserving all the point-group symmetries, the finite sample can either be in a ``square cut" or in a ``diamond cut," with edges at 45$^{\circ}$ relative to each other. $B_{1u}$ pairings and $B_{2u}$ pairings leave certain surfaces gapless. These gapless surfaces are protected by mirror symmetries.  Naturally, a finite sample hosts HOMAs only when the surfaces are gapped, which we summarize in Table \ref{table:0}.

We further show that the BdG Dirac points host an additional $\mathbb{Z}$ monopole charge protected by the $\mathcal{C}_4$ symmetry, which are the same at the same side of the Brillouin zone. As a result, the BdG Dirac points at the north and south poles cannot annihilate each other, and the corresponding HOMAs at the hinges cannot be removed, at least in the weak-pairing limit.

We note that a method based on inversion-symmetry indicator has been proposed in Refs.~\cite{HOTDSC,Indicator_TSC_I,Indicator_TSC}. However, directly applying their results to our case would yield a trivial result. This is because the indicator is only a diagnostic for HOMAs at $k_z=0$, while in our cases HOMAs occur at finite $k_z$'s. Therefore, our results enrich the classification of HOMAs in Ref. \cite{HOTDSC}. 

Based on the same method, we also analyze the topological properties of other pairing channels, out of which we  identify $B_{1g}$ and $B_{2g}$ states as hosting the same monopole charges as the $B_{1u}$ and $B_{2u}$ states.  However, these states host cages of nodal lines in the spectra, thus, the higher-order topology is not well defined. However, each nodal cage still hosts a well-defined $\mathbb{Z}$ monopole charge of two, ensuring the stability of the structure.

The rest of this paper is organized as follows. In Sec. \ref{sec:2} we present the 3D model for a $\mathcal{C}_4$-symmetric Dirac semimetal and the topological protections of bulk Dirac nodes. We perform a real-space Wannier orbital analysis on the 2D slice Hamiltonian between the two bulk Dirac points in Sec. \ref{sec:2a} and find an inevitable filling anomaly. We claim in Sec. \ref{sec:2b} that surface states are gapped out by SOC and HOFAs connecting the projections of bulk Dirac points are present in the 3D rod Hamiltonian. {In Sec. \ref{sec:3}, we focus on the odd-parity pairing states in $B_{1u}$/$B_{2u}$ channels. We show that they are the leading instabilities induced by Ising ferromagnetic fluctuations, shown in Sec. \ref{sec:3a}. In Sec. \ref{sec:3b}, a higher-order $\mathbb{Z}_2$ invariant from the $\mathcal{C}_4$ symmetry is defined by applying Wannier orbital analysis to the 2D subsystem of the BdG Hamiltonian, which characterizes the higher-order corner Majorana zero modes. An additional $\mathbb{Z}$-valued topological invariant protecting HOMAs is identified in Sec. \ref{sec:3d}.  In Sec \ref{sec:3c}, we elucidate the relation between HOMAs with gapless surface states protected by mirror symmetries.  We discuss the effects of  $\mathcal{C}_4$, time-reversal, and inversion-breaking perturbations in Sec. \ref{sec:3e}. We briefly discuss other pairing channels in Sec. \ref{sec:4}. The results are summarized in Sec \ref{sec:5}.}

\section{Dirac semimetals protected by $\mathcal{C}_4$ rotation symmetry}
\label{sec:2}
An effective Hamiltonian of a well-known Dirac semimetal such as Cd$_3$As$_2$ is given by \cite{SatoSC} 
\begin{align}
    H_{0}(\bf k) = &[M_0 - t_{xy}(\cos k_x + \cos k_y) - t_z \cos k_z ]\sigma_z s_0 \nonumber\\
    &+ \eta \sin k_x \sigma_x s_z
     -\eta \sin k_y \sigma_y s_0  \nonumber\\
    &+ \lambda_1 \sin k_z (\cos k_y - \cos k_x)\sigma_x s_x \nonumber\\
    &+ \lambda_2  \sin k_z \sin k_x \sin k_y \sigma_x s_y,
    \label{eq:1}
\end{align}
where $M_0, t_{xy}, t_z, \eta, \lambda_1$ and $\lambda_2$ are model parameters and $\sigma_i$ ($s_i$) denotes Pauli matrices in orbital (spin) spaces.   A pair of Dirac points appear at $(0,0,\pm k_0)$ in the 3D Brillouin zone (BZ) for $t_z > (M_0 - 2t_{xy}) > 0$ and $k_0$ is defined by $M_0 = t_z \cos k_0 + 2t_{xy}$ with $k_0 > 0$. In particular, $\lambda_1$ and $\lambda_2$ are two lowest-order spin-orbit coupling terms allowed by symmetry. {A momentum-dependent chemical potential $\mu({\bf k})\sigma_0 s_0$ is not included in Eq. \eqref{eq:1} since it does not affect the topology.}

The Hamiltonian preserves inversion $\mathcal{I}$, time-reversal $\mathcal{T}$,  three mirror $\mathcal{M}_{x,y,z}$, and a $\mathcal{C}_{4}$ rotation symmetry about the $z$ axis, satisfying
\begin{align}
    \mathcal{I} H_{0}(k_x,k_y,k_z) \mathcal{I}^{-1} &= H_{0}(-k_x,-k_y,-k_z), \nonumber \\
    \mathcal{T} H_{0}(k_x,k_y,k_z) \mathcal{T}^{-1} &= H_{0}(-k_x,-k_y,-k_z), \nonumber \\
    \mathcal{C}_4 H_{0}(k_x,k_y,k_z) \mathcal{C}_4^{-1} &= H_{0}(k_y,-k_x,k_z), \nonumber \\
    \mathcal{M}_x H_{0}(k_x,k_y,k_z) \mathcal{M}_x^{-1} &= H_{0}(-k_x,k_y,k_z), \nonumber \\
    \mathcal{M}_y H_{0}(k_x,k_y,k_z) \mathcal{M}_y^{-1} &= H_{0}(k_x,-k_y,k_z), \nonumber \\
    \mathcal{M}_z H_{0}(k_x,k_y,k_z) \mathcal{M}_z^{-1} &= H_{0}(k_x,k_y,-k_z),
\end{align}
where \begin{align}
&\mathcal{I} = \sigma_z,~~ \mathcal{T} = i s_y \mathcal{K},~~
 \mathcal{C}_4 = i s_z e^{i(\pi/4)\sigma_z s_z},\nonumber\\
 & \mathcal{M}_x = is_x, ~~ \mathcal{M}_y = i\sigma_z s_y, ~~\mathcal{M}_z = is_z.
  \end{align} 
The forms of the symmetry operators are chosen according to the transformation properties of four spin-orbit coupled degrees of freedom with angular momenta $J = \pm 1/2$ and $J = \pm 3/2$. The corresponding point group of $H_{0}(\bf k)$ is $D_{4h}$. 

To see how $\mathcal{C}_4$, $\mathcal{T}$, and $\mathcal{I}$ together protects the Dirac point, we note that $(\mathcal{TI})^2=-1$, which protects a twofold Kramers degeneracy at any point in the BZ. Along $k_x=k_y=0$, every Bloch state can be labeled by its angular momentum under the fourfold rotation symmetry, forming two Kramers pairs. The Dirac points occur at $k_x=k_y=0$ when the $|J|=1/2$ bands and $|J|=3/2$ bands cross, which are robust against any symmetry-preserving perturbations. Here it is important that the two {pairs} of bands carry different absolute values of angular momenta.

One can define a $\mathbb{Z}_2$ monopole charge for the Dirac point at $(0,0,k_0)$~\cite{SatoSC,DSM_BJY_charge,DSM_BJY_classify}, given by
\begin{align}
    \Delta\nu(k_0) =  \[\nu_J (k_0^{+}) - \nu_J (k_0^{-})\] \mod 2,
    \label{eq:4}
\end{align}
where
\begin{align}
 \nu_J (k_0^{\pm})=\[N_J(0,0,k_0^{\pm})- N_J (\pi,\pi, k_0^{\pm})\] \mod 2
 \end{align} 
 and $N_J (0,0, k_0^{\pm})$ is the number of occupied bands with $\mathcal{C}_4$ eigenvalue $J$ at the $\mathcal{C}_4$-invariant momenta $(0,0,k_0^{\pm})$ above (below) $k_0$ along the $\mathcal{C}_4$ invariant axis. 
Time-reversal symmetry and filling constraints ensure that $\nu_J$ is independent of the particular $J$ one chooses, and in our model, it is straightforward to verify that $\Delta\nu(\pm k_0)= 1$.

We will show in the following subsections that the charge defined in Eq. (\ref{eq:4}) is a higher-order topological invariant connected with the Wannier representation for each $k_z$ layer.

\subsection{Wannier representation}
\label{sec:2a}
Consider the 2D subsystem of $H_0(\bf k)$ at a fixed $k_z$ between two Dirac points:
\begin{align}
    H_{\text{2D}}(k_x,k_y) =& [M_0' - (\cos k_x + \cos k_y)]\sigma_z s_0\nonumber\\
    & + \sin k_x \sigma_x s_z - \sin k_y \sigma_y s_0  \nonumber\\
    &+ \lambda_1' (\cos k_y - \cos k_x)\sigma_x s_x \nonumber\\
    &+ \lambda_2'  \sin k_x \sin k_y \sigma_x s_y,
    \label{eq:6}
\end{align}
where $M_0' = M_0 - \cos k_z$, $\lambda_{1,2}'=\lambda_{1,2}\sin k_z$, and we set $t_{xy} = t_z = \eta = 1$ without loss of generality. When $k_z \neq k_0$, $H_{\text{2D}}(k_x,k_y)$ describes a 2D subsystem of a crystalline insulator. Note that the slice Hamiltonian $H_{\text{2D}}(k_x,k_y)$ preserves a 2D time-reversal symmetry defined by the 3D time-reversal operator $\mathcal{T}$ times the mirror reflection $\mathcal{M}_z$, such that 
\begin{align}
    \bar{\mathcal{T}} H_{\text{2D}}(k_x, k_y) \bar{\mathcal{T}}^{-1} = H_{\text{2D}}(-k_x, -k_y),
\end{align}
where $\bar{\mathcal{T}} = \mathcal{TM}_z = i s_x \mathcal{K}$ and $\bar{\mathcal{T}}^2 = +1$. Such a time-reversal symmetry does not enable a Kramers degeneracy and the  edges (corresponding to side surfaces at a given $k_z$) are in general gapped. Lacking protected edge states, $H_{\text{2D}}(k_x,k_y)$ for $k_z\neq \pm k_0$ can be represented by Wannier states.

Constrained by spatial symmetries and $\mathcal{TI}$ symmetry, there are two possible representations of two filled Wannier states, with {a pair of Wannier centers both at Wyckoff position ${\bf r} = (1/2,1/2)$ or both at ${\bf r} = (0,0)$.} The configuration corresponding to $H_{\text{2D}}(k_x,k_y)$ can be obtained by investigating the $\mathcal{C}_4$ eigenvalues of the occupied bands at $\mathcal{C}_4$ invariant momentum $\Gamma = (0,0)$ and $M = (\pi,\pi)$ in the 2D Brillouin zone, which we denote as $C_{\Gamma} e^{\pm i \frac{\pi}{4}}$ and $C_{M} e^{\pm i \frac{\pi}{4}}$. Due to $\mathcal{TI}$ symmetry ensuring Kramers doublets, $C_M$ and $C_\Gamma$ are both real numbers --- for $|J| = 1/2$, the corresponding $C_{M,\Gamma} = 1$ and for $|J| = 3/2$, $C_{M,\Gamma} = -1$.

\begin{figure}
	\includegraphics[width=1\columnwidth]{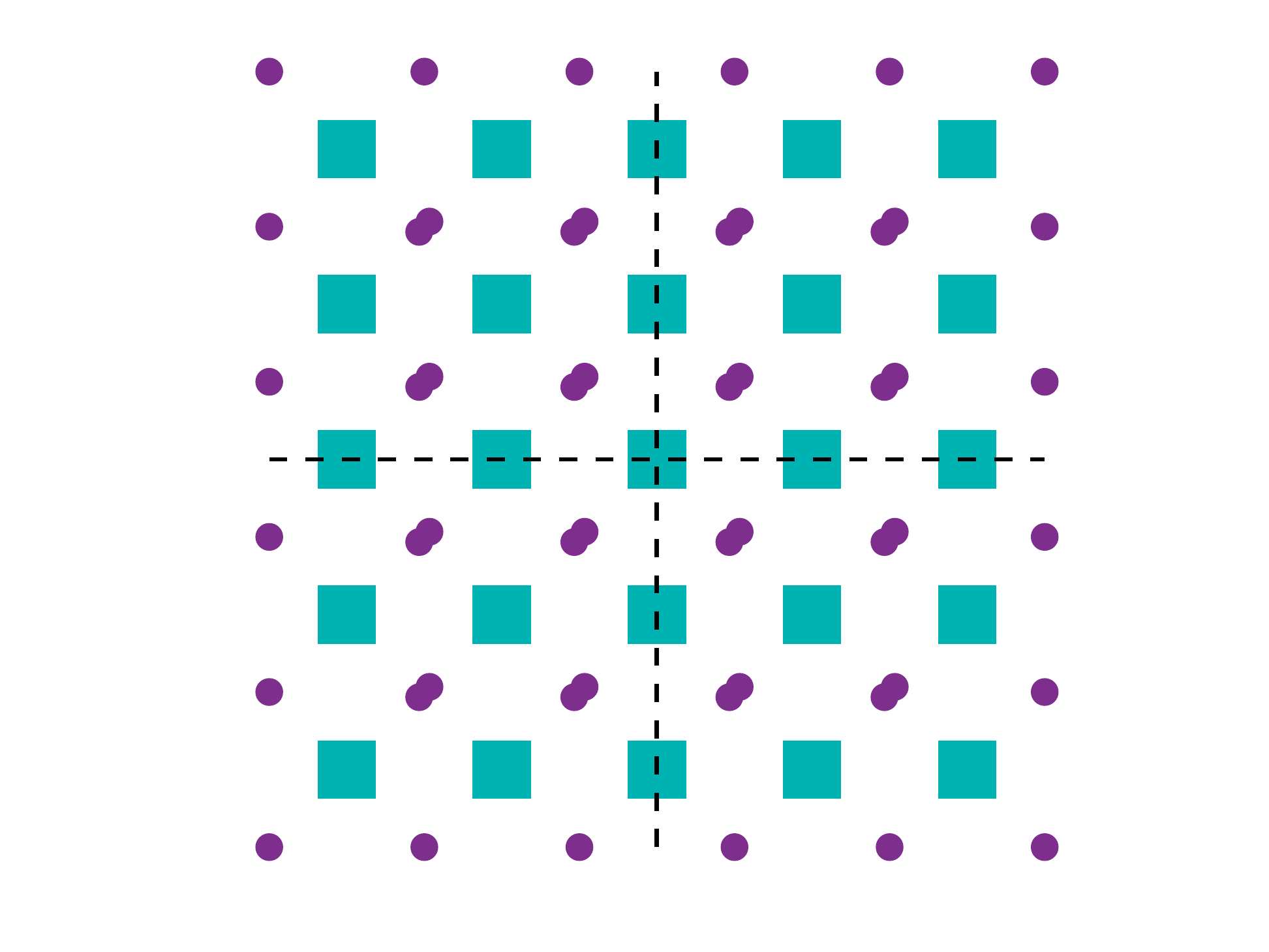}
   \caption{Wannier representations for the nontrivial phase in 2D subsystem of the Hamiltonian in Eq. ($\ref{eq:6}$). Cyan squares represent atom sites and purple dots denote Wannier orbital centers at ${\bf r} = (\frac{1}{2},\frac{1}{2})$. Two dashed lines denote mirror planes and their crossing point is the rotation axis. The corner atom site in the finite sample acquires a charge $\pm\frac{1}{2}e$ due to the filling anomaly.} 
\label{fig:normal}
\end{figure}

One can define a topological invariant
\begin{align}
(-1)^{\nu} = C_M C_\Gamma,
\label{eq:8}
\end{align}
{such that $\nu = 1$ represents the configuration of two Wannier orbitals centered at the Wyckoff position ${\bf r} = (1/2,1/2)$ and $\nu = 0$ corresponds to the trivial case when the Wannier centers are at ${\bf r} = (0,0)$ \cite{TiwariJahinWang} (details in Appendix. \ref{appen:0}). Under  open boundary conditions, the nontrivial case $\nu = 1$ is sketched in Fig. \ref{fig:normal} and exhibits a filling anomaly, i.e., it is impossible for the system to maintain both charge neutrality and $\mathcal{C}_4$ rotation symmetry due to the fact that Wannier centers are shifted from the atomic positions, leading to an obstructed atomic limit \cite{TQChemistry}. It is separated by a gap closing point to its trivial phase where Wannier centers coincide with ionic sites. The 2D subsystem is in a second-order topological insulating phase \cite{Fang_fillinganomaly,SongHOTI,Hughes_corner,BBH_Science,Hughes_corner_PRB,frac_corner_SOC}.} Even though we have only depicted  a ``square cut" geometry, it can be readily verified that the filling anomaly exists for \emph{any} finite geometry that preserves the $\mathcal{C}_4$ symmetry, including a ``diamond cut" along the diagonal direction.

From Eq.~\eqref{eq:6}, for $-2 < M_0-\cos k_z < 2$, which is precisely the range of $k_z$ between the two Dirac points, the filled bands carries $|J|=3/2$ at $\Gamma$ and $|J|=1/2$ at $M$. Therefore for this range of $k_z$, we have $\nu=1$, exhibiting a filling anomaly.

Note that the second-order topological invariant $\nu$ defined here is equivalent with $\nu_J$ defined in Eq.~\eqref{eq:4}, and the topological charge $\Delta\nu$ of the Dirac point can be interpreted as a relative $\mathbb{Z}_2$ invariant of two $k$-space layers sandwiching the Dirac point.

\subsection{HOFAs}
\label{sec:2b}
Corresponding to a nontrivial $\nu=1$, one can show that the system in a rod geometry hosts gapless HOFAs on the vertical hinges, provided that the side surfaces are gapped \cite{sub_topology}. 
We note that this necessarily excludes the $k_z=0$ slice of the side surfaces due to a mirror Chern number~\cite{SatoSC}. However, surface states with momenta  $k_z\neq 0,\pm k_0$ can all be gapped out without breaking the symmetries. Our findings are consistent with those obtained in Refs.~\cite{NC_HOFA,Fang_classifyHOFA,Fang_fillinganomaly}, and we present our full analysis here for completeness.

Indeed, this can be seen from splitting Eq.~\eqref{eq:6} in two parts:
\begin{align}
H_{\rm 2D} \equiv & H_0+\lambda_1' (\cos k_y - \cos k_x) \sigma_x s_x  \nonumber\\
&+ \lambda_2'  \sin k_x \sin k_y \sigma_x s_y,
\label{eq:split}
\end{align}
where 
\begin{align}
    H_0=& [M_0' - (\cos k_x + \cos k_y)]\sigma_z s_0\nonumber\\
    & + \sin k_x \sigma_x s_z - \sin k_y \sigma_y s_0.   \label{eq:6'}
\end{align}
It can be readily verified that $H_0$ hosts helical surface states. On the $yz$ surfaces they satisfy 
\begin{align}
\sigma_y  s_z  |\Psi\rangle = \pm |\Psi\rangle.
\end{align}
For small $k_y$, the Hilbert space of the surface Hamiltonian is spanned by $|\sigma_y = 1\rangle\otimes|s_z = 1\rangle$ and $|\sigma_y = -1\rangle\otimes|s_z = -1\rangle$, giving rise to
\begin{align}
H_{0}^{yz} \simeq k_y s_z.
\label{eq:helical1}
\end{align}
The surface dispersion {$H_{0}^{xz}$} on the $xz$ surface can be similarly obtained. 
It  then follows that $\lambda_1'$ term in \eqref
{eq:split} gets projected onto the surface {and becomes $\lambda_1' \langle k_x^2\rangle s_y$, which gaps out the surface Hamiltonian \cite{DSMnoFA2,DSMnoFA3,ZhangDSM}.} To see how hinge modes emerge, it is helpful to set $\lambda_2'=0$, and consider a smooth hinge where $xz$ and $yz$ surfaces are smoothly connected. On this smooth manifold, due to the $\cos k_y - \cos k_x$ dependence, $\lambda_1'$ term necessarily vanishes in the diagonal direction, indicating a mass domain wall of the helical fermion in Eq. \eqref{eq:helical1}, therefore leading to a localized zero mode at the hinge. The $\lambda_2'$ term has a nonzero projection at the corner, i.e., the diagonal direction, and shifts the energy of the hinge mode. Nonetheless, due to the $\mathcal{C}_4$ symmetry, such an energy shift is the same for all four corners, and thus the hinge modes are degenerate.

The same argument can be applied to obtain hinge states in a ``diamond cut," in which the relevant surfaces are $(x+y),z$ and $(x-y),z$. Without $\lambda_{1,2}'$ terms, the gapless edge states are given by, e.g.,
\begin{align}
H_{0}^{(x-y)z}\simeq (k_x-k_y) \(\sigma_xs_z + \sigma_ys_0\)
\end{align}
and
\begin{align}
\frac{\sigma_x+\sigma_y s_z}{\sqrt{2}}|\Psi\rangle = \pm |\Psi\rangle.
\end{align} 
The $\lambda_2'$ term gaps out the surface states such that when viewed as a smooth edge, the gap changes sign from the $(x-y),z$ surface to the $(x+y),z$ surface, leading to a zero mode. Further, the $\lambda_1'$ term displaces the bound state energies while maintaining their degeneracy for all corners.

\begin{figure}
	\subfloat[]{\includegraphics[width=0.8\columnwidth]{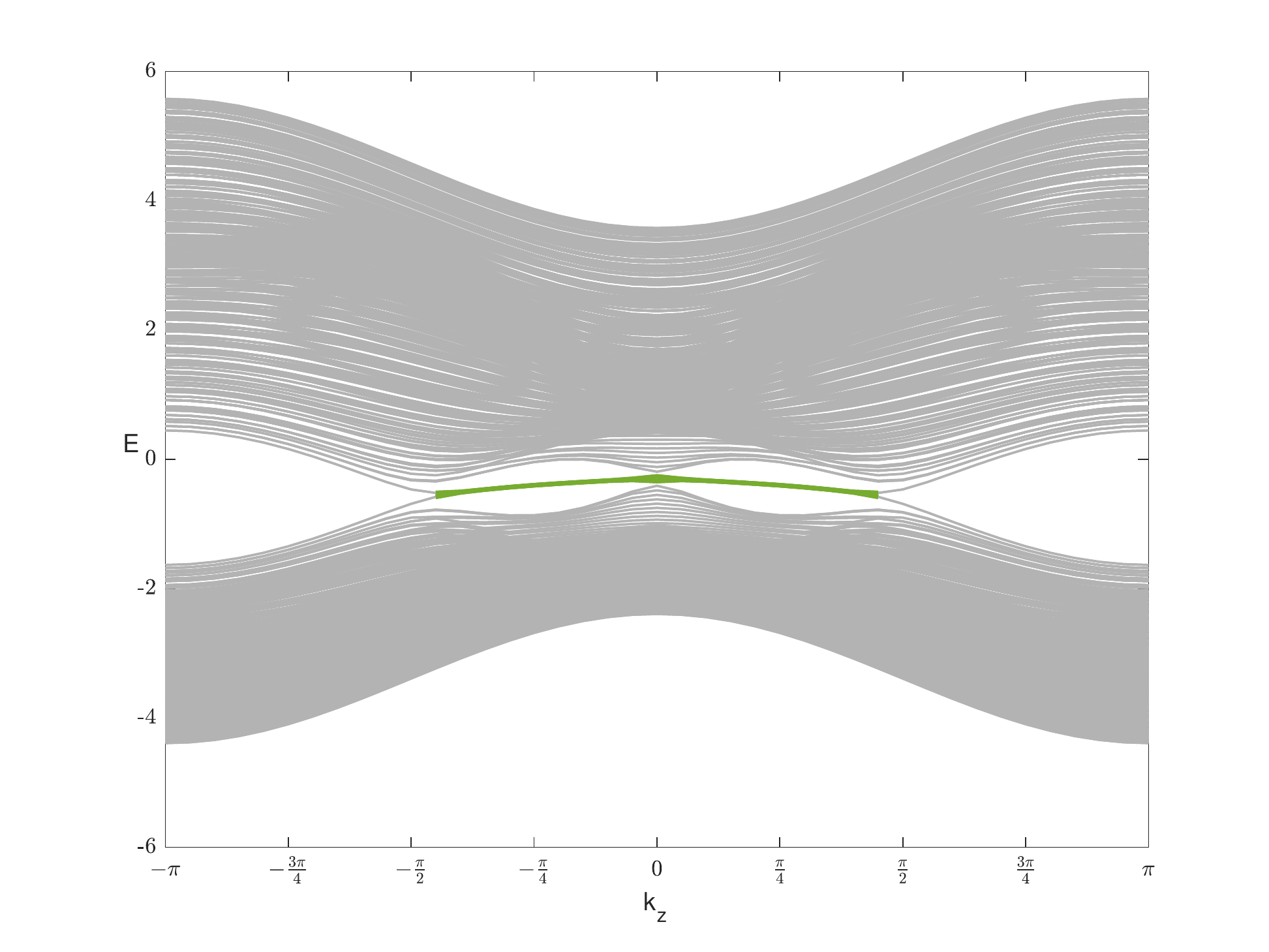}}
	\hfill
	\subfloat[]{\includegraphics[width=0.8\columnwidth]{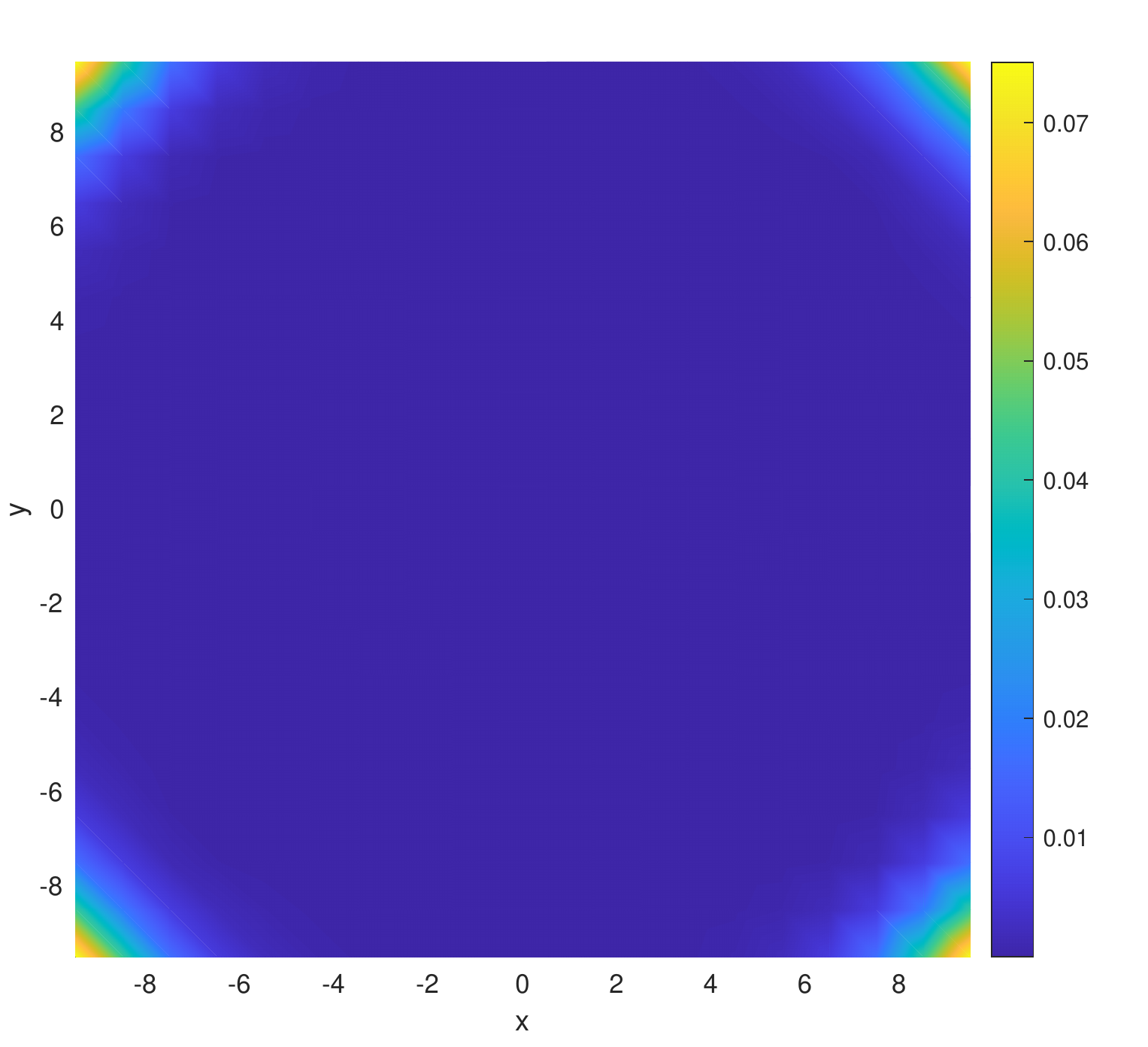}}
   \caption{(a) Energy spectrum of the rod Hamiltonian $H_0(x,y,k_z)$ as a function of $k_z$ for $N = 20$. {A momentum-dependent term $-0.3\(\cos k_x + \cos k_y\)\sigma_0 s_0$ is added in the numerics to lift the ``accidental" particle-hole symmetry.} Hinge states are marked green. (b) Wave-function distributions of the mid-gap states on a 2D slice in (a) at $k_z = \pi/4$. Parameters are $M_0 = 2$ and $t_{xy} = t_z = \eta = \lambda_1 = \lambda_2 = 1$. The corner states are visible.} 
\label{fig:2}
\end{figure}

\begin{figure}
    \includegraphics[width = 2in]{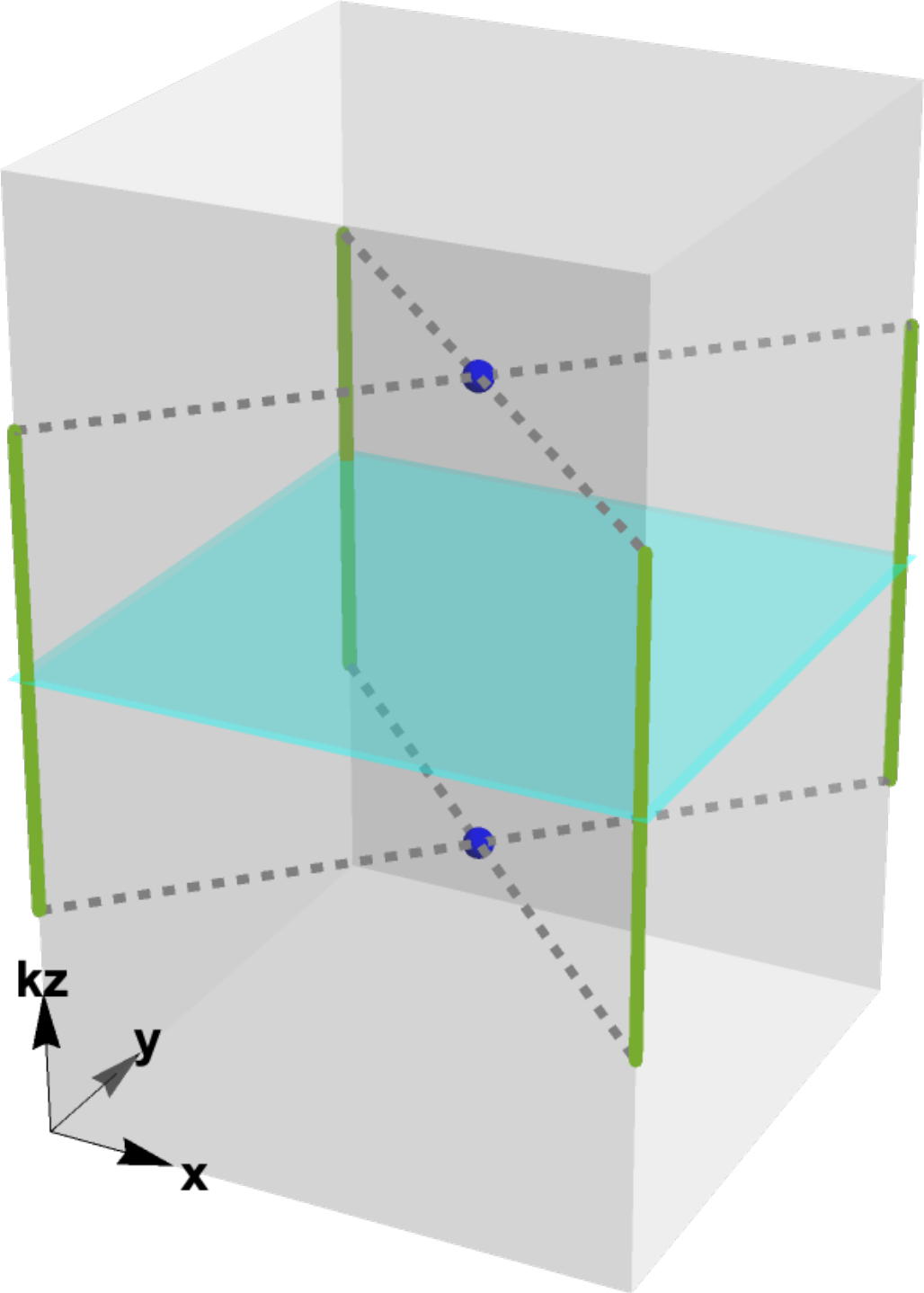}
    \caption{The $\mathcal{C}_4$-symmetric Dirac semimetal phase in a rod geometry. Two blue dots denote the bulk Dirac points and the cyan plane corresponds to a nontrivial mirror Chern number at $k_z = 0$. HOFAs (green) cross $k_z = 0$ and terminate at the projections of bulk Dirac nodes onto four vertical hinges.}
    \label{fig:3}
\end{figure}

In Fig.~\ref{fig:2} we show numerical results confirming the existence of higher-order corner states in a 2D square cut subsystem. We schematically plot the HOFAs in the rod Hamiltonian $H_0(x,y,k_z)$ in Fig.~\ref{fig:3}.

\section{$B_{1u}$ and $B_{2u}$ pairing in Dirac Semimetals}
\label{sec:3}
\begin{table*}
\begin{tabular}{c|c|c|c|c|c|c}
    \hline
    \hline
        Representative pairing order& Irrep & $\mathcal{C}_4 = is_z e^{i\frac{\pi}{4}\sigma_z s_z}$ & $\mathcal{I} = \sigma_z$ & $\mathcal{M}_x = is_x$ & $\mathcal{M}_y = i\sigma_z s_y$ & $\mathcal{M}_z = is_z$\\
    \hline
    $\Delta_1 \sigma_y$  & $B_{1u}$  & $-$    & $-$ &   $-$ & $-$ & $-$\\
    $\Delta_2 \sigma_y s_z$& $B_{2u}$  &   $-$  & $-$  & $+$ & $+$ & $-$\\
    \hline
    \hline
   \end{tabular}
\caption{Transformation properties of $\Delta_1$ and $\Delta_2$, which forms $B_{1u}$ and $B_{2u}$ irreps of $D_{4h}$.}
\label{table:1}
\end{table*}
With a nonzero chemical potential, there exist two closed Fermi surfaces near the two Dirac points. 
In this section we focus on the odd-parity pairing states belonging to $B_{1u}$ and $B_{2u}$ irreducible representations. They are described by the following equal-spin pairing terms:
\begin{align}
B_{1u}:~~& \Delta_1  \psi^\dagger({\bf k}) (\sigma_y s_0)  \psi^\dagger({-\bf k}),\nonumber\\
B_{2u}:~~& \Delta_2  \psi^\dagger({\bf k}) (\sigma_y s_z)  \psi^\dagger({-\bf k}),
\label{eq:B1uB2u}
\end{align}
and we list their transformation properties under point-group symmetries of $D_{4h}$ in Table~\ref{table:1}. Note that, due to the particle-particle nature, the pairing order transforms under symmetry operations as, e.g., 
\begin{align}
\Delta_1 \sigma_y s_0 \to \Delta_1 \mathcal{M}_x \sigma_y s_0  \mathcal{M}_x^T = -\Delta_1 \sigma_y s_0.
\end{align}

These pairing channels are favored by certain electronic interactions, for example, ferromagnetic spin fluctuations. In fact, we show below that in the presence of short-ranged Ising ferromagnetic fluctuations, the leading pairing instabilities are in these two channels. We will then analyze the topological properties of these two pairing states.

\subsection{$B_{1u}$ vs $B_{2u}$ instabilities}
\label{sec:3a}
Let us consider an electronic interaction that is attractive in both $B_{1u}$ and $B_{2u}$ channels. For concreteness, we focus on interactions mediated by short-ranged Ising ferromagnetic fluctuations, given by in second-quantized form
\begin{align}
\mathcal{H}_{\rm fm} = V\int d{\bf r} \[\psi^\dagger ({\bf r}) (\sigma_0 s_z) \psi ({\bf r}) \]^2,
\end{align}
where $V<0$ is the strength of the ferromagnetic fluctuation. It is well-known that ferromagnetic spin fluctuations suppress spin-singlet $s$-wave ($A_{1g}$) pairing and promote odd-parity pairing \cite{BJY_spin_polarized, ferro_triplet,near_ferro_triplet}. It turns out the leading instabilities with this interaction are  in the $B_{1u}$ and $B_{2u}$ channels that are equal-spin, orbital-singlet, and constant in $\bf k$. The linearized gap equations are given by
\begin{align}
\Delta_1 \sigma_y &= -T\sum_k V s_z G(k) \(\Delta_1 \sigma_y\) G^{\rm T}(-k)s_z,\\
\Delta_2 \sigma_ys_z &= -T\sum_k V s_z G(k) \(\Delta_2 \sigma_y s_z\) G^{\rm T}(-k)s_z,
\end{align}
where $k\equiv (i\omega_m,{\bf k}), G(k)=[i\omega_m - H_0({\bf k})+\mu]^{-1}$ is the (matrix) fermionic Green's function and $\omega_m$ is the fermionic Matsubara frequency. It can be straightforwardly verified that $\Delta_{1}\sigma_y$ and $\Delta_2\sigma_ys_z$ are indeed eigenvectors of the kernel.

After multiplying both sides respectively by $\sigma_y$ and $\sigma_ys_z$ and taking the trace, we have 
\begin{align}
-1/V=&\ T_{c1}\sum_k \text{Tr}\[\sigma_y G(k)\sigma_y G^{\text{T}}(-k)\]\nonumber\\
 \equiv & \sum_{FS_{\pm}}F_1({\bf k})\frac{\tanh \frac{\epsilon({\bf k})-\mu}{2T_{c1}}}{\epsilon({\bf k})-\mu},
 \label{eq:gap1}
 \end{align}
 \begin{align}
-1/V=&\ T_{c2}\sum_k \text{Tr}\[\sigma_ys_z G(k)\sigma_ys_z G^{\text{T}}(-k)\]\nonumber \\
\equiv & \sum_{FS_{\pm}}F_2({\bf k})\frac{\tanh \frac{\epsilon({\bf k})-\mu}{2T_{c2}}}{\epsilon({\bf k})-\mu},
\label{eq:gap2}
\end{align} 
where $\epsilon({\bf k})$ is the dispersion of $H_0({\bf k})$ and we have assumed $\mu > 0$. Note that the momentum summation $FS_{\pm}$ indicates that only electrons near two Fermi surfaces surrounding $(0,0,\pm k_0)$ form Cooper pairs. Beyond the specific interaction we consider, we note that Eqs.~\eqref{eq:gap1} and \eqref{eq:gap2} hold for \emph{any} short-ranged interaction attractive in these channels. For example, these pairing channels can also be induced by  inter-orbital attractive interactions which dominate over intra-orbital ones \cite{SatoSC,SC_dopedDSM}.

The derivations of form factors $F_{1,2}({\bf k})$ are given in Appendix \ref{appen:1}, where we find that
\begin{align}
	F_1({\bf k}) & \simeq \frac{k_x^2+k_y^2 + \lambda_1^2\(\frac{k_x^2-k_y^2}{2}\)^2\sin^2 k_0}{\mu^2},\\
	F_2({\bf k}) & \simeq \frac{k_x^2+k_y^2 + \lambda_2^2k_x^2k_y^2\sin^2 k_0}{\mu^2}.
\end{align}
Here in $F_{1,2}({\bf k})$ we have  set $\epsilon(\bf k)=\mu$ and hence the $\simeq$ sign. For approximately spherical Fermi surfaces, we find that $\sum_{\bf k}F_1 ({\bf k}) > \sum_{\bf k}F_2 ({\bf k})$, i.e., $T_{c1} > T_{c2}$ when $|\lambda_1| >|\lambda_2|$,  thus favoring $\Delta_1$ as the pairing state, and vice versa. This conclusion holds for small Fermi surfaces that can be sufficiently approximated by a $k\cdot p$ Hamiltonian. For larger Fermi surfaces a more detailed analysis of the realistic band structure is needed.

We note that $F_{1,2}({\bf k})$ vanishes at $k_x=k_y=0$, which are the polar points on the two disjoint Fermi surfaces. These $\bf k$ points thus do not participate in pairing.  In fact, as we shall see, there exist gapless nodal points at the poles of the Fermi surfaces.

\subsection{Higher-order topology and HOMAs}
\label{sec:3b}
\begin{figure}
	\subfloat[$H_{\text{BdG}}(\Gamma)$]{\includegraphics[width=1\columnwidth]{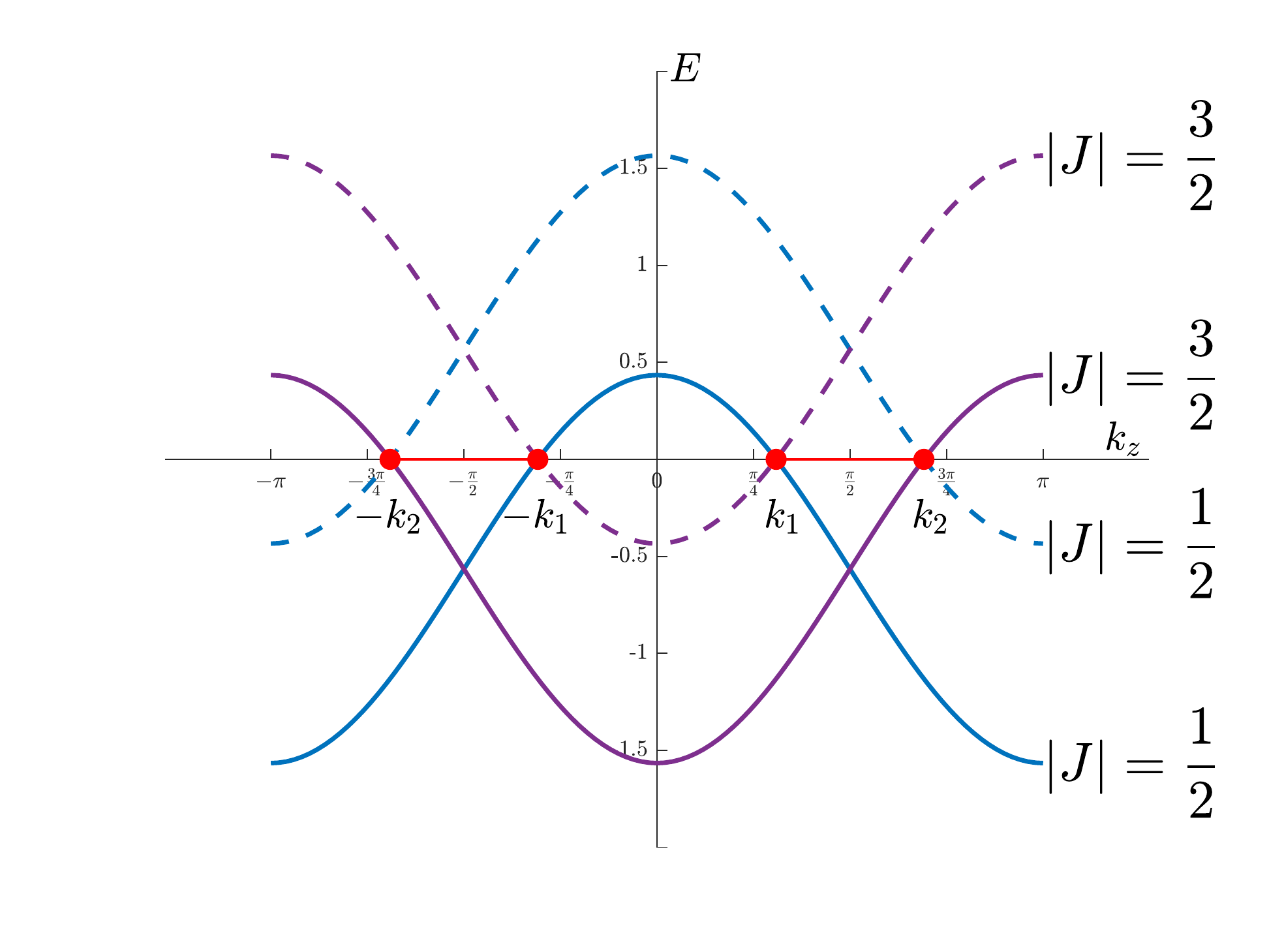}}
	\hfill
	\subfloat[$H_{\text{BdG}}(M)$]{\includegraphics[width=1\columnwidth]{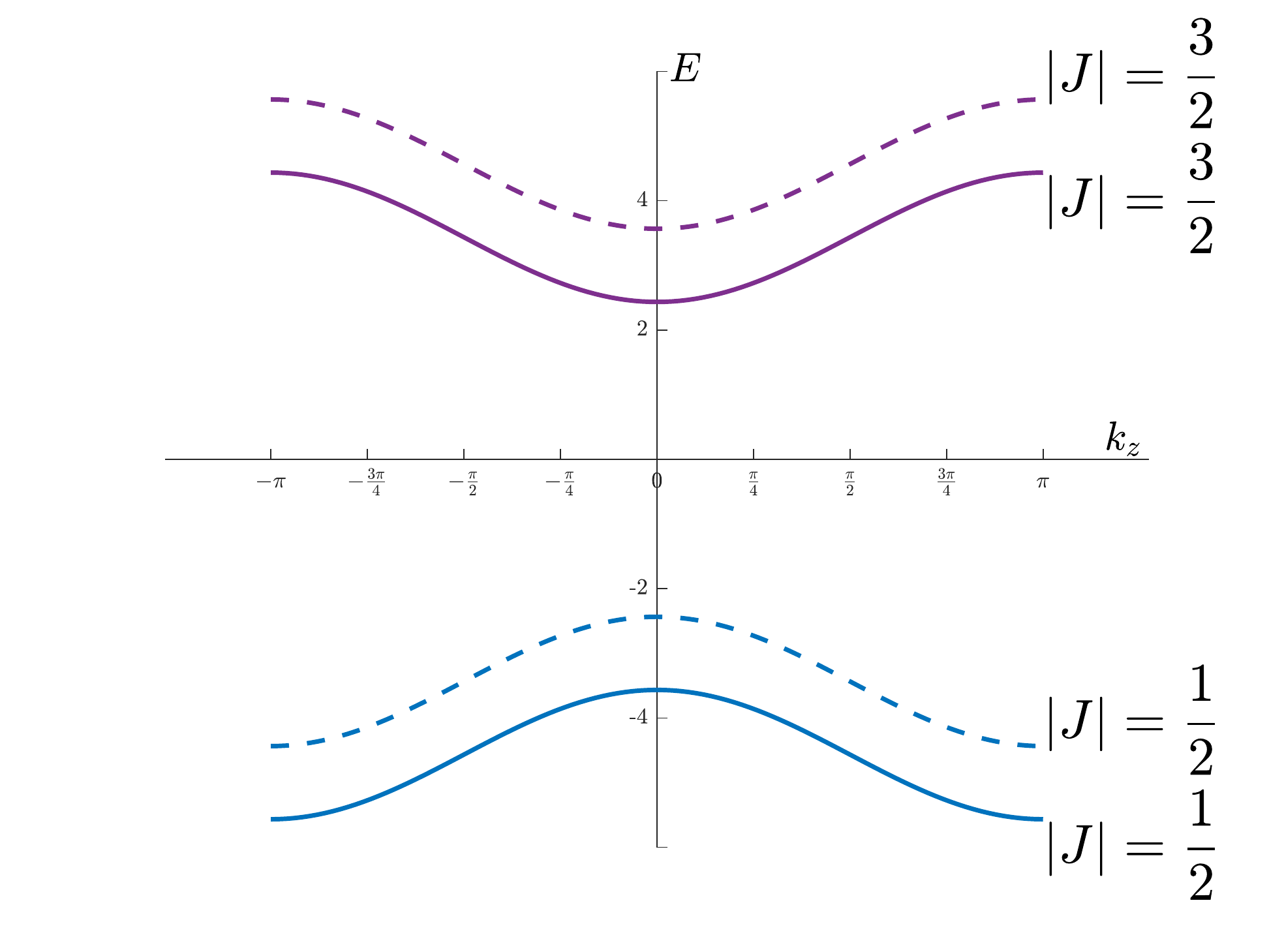}}
   \caption{Energy spectrum of (a) $H_0(0,0,k_z)$ and (b) $H_0(\pi,\pi,k_z)$ with $M_0 = 2$ and $\mu = \Delta_1 = 0.4$. All bands are two-fold degenerate due to the product of time-reversal and inversion symmetries $(\bar{\mathcal{T}}\tilde{\mathcal{I}})^2 = -1$. Purple bands correspond to $|J| = 3/2$ while blues bands correspond to $|J| = 1/2$. Dashed lines denote BdG shadow bands. In (a), there are four BdG Dirac points (marked by red dots) at $k_z = \pm k_1$ and $\pm k_2$, between which the 2D slice Hamiltonian represents a nontrivial higher-order topological superconducting phase.}
\label{fig:5}
\end{figure}

The corresponding Bogoliubov-de Gennes (BdG) Hamiltonian for the $B_{1u}$ ($B_{2u}$) pairing state is 
\begin{align}
    H_{\text{BdG}}({\bf k}) = & (M_0 - \cos k_x - \cos k_y - \cos k_z) \sigma_z\tau_z\nonumber\\
    	& + \sin k_x \sigma_x s_z - \sin k_y \sigma_y\tau_z -\mu \tau_z\nonumber\\
                     & + \lambda_1\sin k_z (\cos k_y - \cos k_x)\sigma_x s_x\nonumber\\
                     & + \lambda_2\sin k_z \sin k_x \sin k_y \sigma_x s_y \tau_z\nonumber\\
                     & + \Delta_1 \sigma_y \tau_y \(+ \Delta_2 \sigma_y s_z\tau_x \),
                     \label{eq:17}
\end{align}
where $\tau_{x/y/z}$ are Pauli matrices in the Nambu space. The symmetry operators are elevated to the BdG level as 
\begin{align}
\tilde{\mathcal{I}} = \sigma_z\tau_z, 
~~\tilde{\mathcal{T}} = i s_y \mathcal{K},
~~\tilde{\mathcal{C}_4} = i s_z e^{i(\pi/4)\sigma_z s_z \tau_z}.
\end{align} Mirror symmetry operators are different for $B_{1u}$ and $B_{2u}$ order, such that
\begin{align}
\textrm{for $B_{1u}, ~~~$}&\tilde{\mathcal{M}}_x = i s_x,~~ \tilde{\mathcal{M}}_y = i \sigma_z s_y \tau_z, \nonumber \\ 
\textrm{for $B_{2u}, ~~~$}&\tilde{\mathcal{M}}_x = i s_x\tau_z,~~\tilde{\mathcal{M}}_y = i\sigma_z s_y.
\end{align}
Since $\Delta_1$ and $\Delta_2$ in general do not coexist from our energetic analysis, all point-group symmetries are preserved. In addition, a particle-hole symmetry $\tilde{\mathcal{P}} = \tau_x \mathcal{K}$ is present, such that $\tilde{\mathcal{P}} H_{\text{BdG}}({\bf k}) \tilde{\mathcal{P}}^{-1} = -H_{\text{BdG}}(-{\bf k})$.

Before we analyze the higher-order topology, it can be directly checked that Eq.~\eqref{eq:17} has a mirror Chern number $C_M=2$ at $k_z=0$. Compared with the normal states, the doubling in $C_M$ comes from the Nambu space. This corresponds to protected gapless surface states at $k_z = 0$. For all other $k_z$ values the 2D subsystem does not have nontrivial first-order topology, but may still possess nontrivial second-order topology.

By directly diagonalizing the Hamiltonian, we see that $\Delta_1$ and $\Delta_2$ pairing terms both create nodal points on the north and south poles of the spherical Fermi surface in the bulk BdG spectra. For the $B_{1u}$ order parameter $\Delta_1$, we write $H_{\text{BdG}}({\bf k})$ in Eq. (\ref{eq:17}) along $(0,0,k_z)$:
\begin{align}
    H_{\text{BdG},\Gamma}(k_z) & = -\mu\tau_z - \cos k_z \sigma_z \tau_z + \Delta_1 \sigma_y \tau_y,
\end{align}
where we set $M_0 = 2$ without loss of generality. The corresponding energy dispersion is
\begin{align}
    E = \pm \cos k_z \pm \sqrt{\mu^2 + \Delta_1^2},
\end{align}
which we plot in Fig. \hyperref[fig:5]{\ref{fig:5}(a)}. We see that there exist four fourfold-degenerate crossing points at zero energy: these are the Dirac points in the BdG spectrum (hereafter referred to as ``BdG Dirac points") pinned to zero energy by particle-hole symmetry. Their locations are at $(0,0,k_z)$ with 
\begin{align}
k_z &= \pm k_1 =\pm \arccos\(\sqrt{\mu^2 + \Delta_1^2}\)\textrm{, and }\nonumber\\
k_z &=\pm k_2 \equiv \pm( \pi - k_1).
\end{align}
In the weak-pairing limit, $\Delta\ll \mu$, and the BdG Dirac points are located at the north and south poles of the two Fermi surfaces.

Using $B_{1u}$ order parameter as an example, (the analysis for the $B_{2u}$ order parameter $\Delta_2$ is exactly the same until Sec.~\ref{sec:3c}), the 2D subsystem of $H_{\text{BdG}}({\bf k})$ for a $k_z$ slice is given by
\begin{align}
    H_{\text{BdG, 2D}}(k_x,k_y) =&   (M_0' - \cos k_x - \cos k_y) \sigma_z\tau_z \nonumber\\
    &+ \sin k_x \sigma_x s_z - \sin k_y \sigma_y\tau_z \nonumber\\
    & + \lambda_1' (\cos k_y - \cos k_x)\sigma_x s_x\nonumber\\
    & + \lambda_2' \sin k_x \sin k_y \sigma_x s_y \tau_z\nonumber\\
    &+ \Delta_1 \sigma_y \tau_y   -\mu \tau_z.
    \label{eq:24}
\end{align} 
$H_{\text{BdG}}(k_x,k_y)$ satisfies the 2D version of time reversal , particle-hole, chiral symmetries and a BdG version of $\tilde{\mathcal{C}}_4 = \text{diag}\{\mathcal{C}_4, -\mathcal{C}_4^*\}$. The system above can be regarded as an insulator before taking the particle-hole symmetry into account, and we can deduce a similar topological invariant following the HOFAs case. The difference is that the corner states in this Hamiltonian are pinned at zero energy due to the particle-hole symmetry, hence, they are Majorana zero modes. Equation (\ref{eq:24}) then describes a second-order topological superconducting phase in 2D.

In Fig.~\ref{fig:5}, we label the angular momenta of each BdG band. We see that for $|k_z| > k_2$ and for $|k_z|<k_1$, the four negative-energy bands of the 2D subsystem in Eq. (\ref{eq:24}) carry the same $\mathcal{C}_4$ eigenvalues both at $\Gamma$ and $M$. For $k_1 < k_z < k_2$ and $-k_2 < k_z < -k_1$, there are two negative-energy bands with $|J| = 1/2$ (i.e., $\mathcal{C}_4 = e^{\pm i\frac{\pi}{4}}$) and the other two negative-energy bands with $|J| = 3/2$ (i.e., $\mathcal{C}_4 = e^{\pm i\frac{3\pi}{4}}$) in $H_{\text{BdG}}(\Gamma)$ --- this is simply derived from the normal state band structure. On the other hand, both pairs of negative-energy bands at $M$ point have $|J| = 1/2$ [see Fig. \hyperref[fig:5]{\ref{fig:5}(b)}].

\begin{figure}
    \includegraphics[width=1\columnwidth]{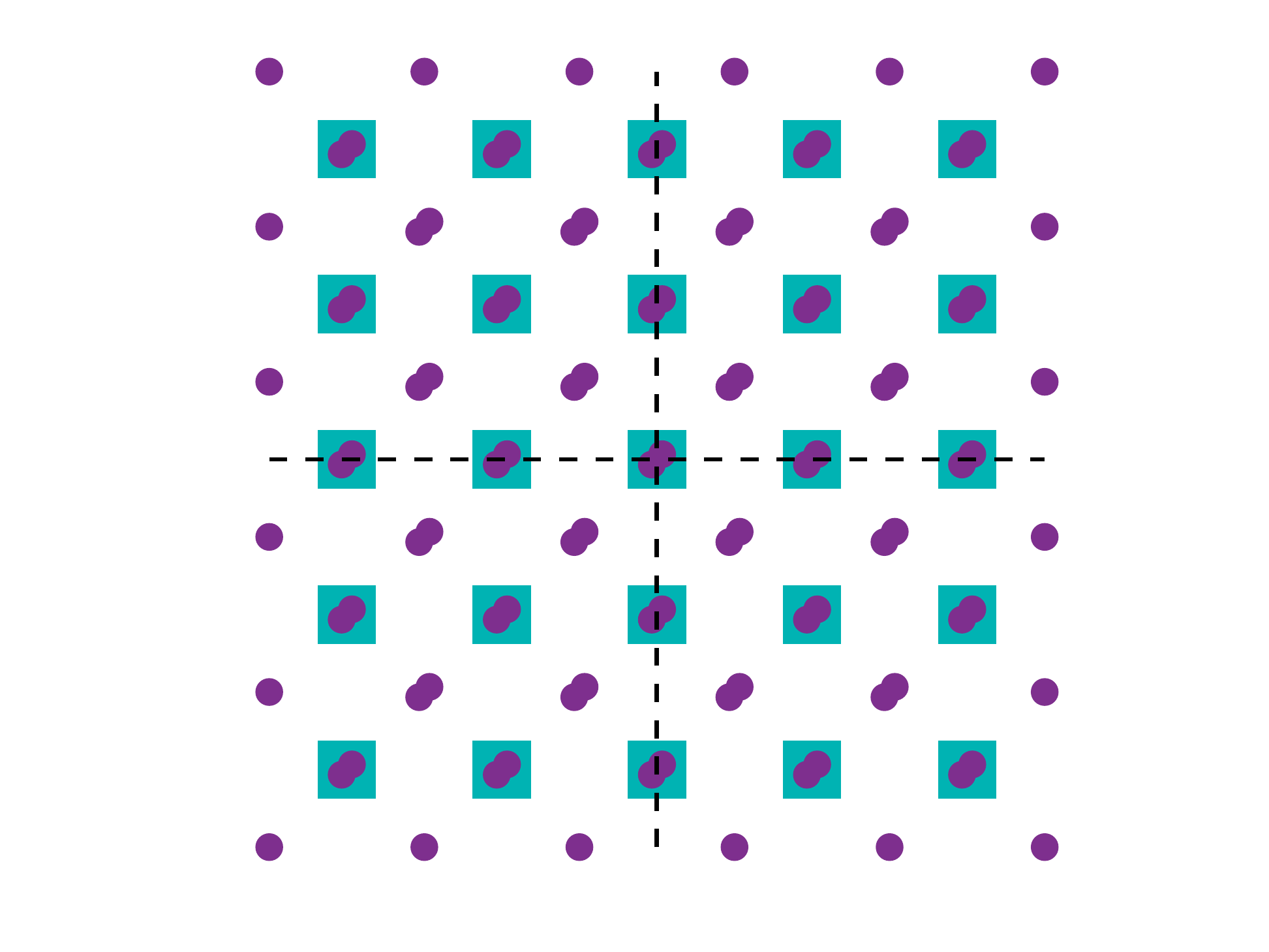}
    \caption{Wannier representations for the nontrivial phase in 2D subsystem of the BdG Hamiltonian in Eq. ($\ref{eq:24}$). Two Wannier orbitals are centered at ${\bf r} = (0,0)$ and the other two at ${\bf r} = (\frac{1}{2},\frac{1}{2})$. Each corner acquires a charge $\pm\frac{1}{2}e$ due to the filling anomaly, indicating Majorana zero modes.}
    \label{fig:4}
\end{figure}

Formally, the same $\mathbb{Z}_2$ topological invariant for the 2D subsystem can be  constructed as a characterization of the second-order topological superconducting phase. With four negative-energy BdG bands, the topological invariant $\nu$ is given by
\begin{align}
    (-1)^{\nu} = \prod_{i} C_{\Gamma,i}C_{M,i},
    \label{eq:28}
\end{align}
where $C_{\Gamma,i}$ is defined the same way as in \eqref{eq:8} and $i$ runs over two sets of twofold-degenerate BdG bands.
It is now straightforward to see that $\nu=1$ for $-k_2 < k_z < -k_1$ and $k_1 < k_z < k_2$, while $\nu=0$ otherwise.  Mapping to real space, the nontrivial 2D system corresponds to the configuration sketched in Fig. \ref{fig:4}, where two Wannier orbitals are centered at ${\bf r} = (1/2,1/2)$ and two Wannier centers are at ${\bf r}= (0,0)$, displaying a filling anomaly. The two $k_z$ slices sandwiching each of the four BdG Dirac points at $(0,0,\pm k_1)$ and $(0,0,\pm k_2)$ differ by $\Delta \nu =1$, which can be viewed as the $\mathbb{Z}_2$ monopole charge of the BdG Dirac points.

To highlight the role of the pairing order in determining the topology, it is helpful to go to the weak-pairing limit $\Delta\ll \mu$, i.e., when the pairing order does not change the band structure away from the Fermi surfaces. Note that at $M$, one of the negative-energy band with {$J = \pm 1/2$} is the same as the normal state, while the other is the BdG shadow of the unfilled bands with {$J =\pm  3/2$}. Importantly, the $B_{1u}/B_{2u}$ pairing order parameter carries an angular momentum $\Delta J = 2$, which causes the shadow band to take a different angular momentum from the original band, which is 
\begin{align}
J= \(\mp \frac 32 +2\) \mod 4 = \pm \frac 12.
\label{eq:angular_momenta}
\end{align}
{The above result comes from the nontrivial anti-commutation relation between the 2D particle-hole symmetry and $\tilde{\cal C}_4$ symmetry
\begin{align}
\left\{{\cal PM}_z, \tilde{\cal C}_4\right\} = 0.
\end{align}
For an arbitrary band $\psi_1$ carrying angular momenta $J_1$ at a $\tilde{\cal C}_4$-invariant point that satisfies $\tilde{\cal C}_4 \psi_1 = e^{i\frac{\pi}{2}J_1}\psi_1$, its BdG shadow band $\({\cal PM}_z\psi_1\)$ carries angular momenta $(2-J_1)$ mod 4 according to $\tilde{\cal C}_4\({\cal PM}_z\psi_1\) = -{\cal PM}_z e^{i\frac{\pi}{2}J_1}\psi_1 = e^{i\frac{\pi}{2}(2-J_1)}\({\cal PM}_z\psi_1\)$, which agrees with Eq. \eqref{eq:angular_momenta}.
}

As at the $\Gamma$ point all the negative-energy states have $J=\pm 1/2, \pm 3/2$, there is a mismatch between $\Gamma$ and $M$ points, which leads to one pair of  Wannier states centered at ${\bf r} = (1/2,1/2)$ in Fig.~\ref{fig:4}. Here we see that the nontrivial pairing symmetry $B_{1u}/B_{2u}$ plays an important role, just as much as the normal state band structure. 
Had the pairing order transformed trivially under $\mathcal{C}_4$, the BdG shadow band would carry instead $J=\pm 3/2$, leading to an on-site Wannier center at ${\bf r} = (0,0)$ and no filling anomaly.

By the same token as in the normal state in a rod geometry, due to the filling anomaly, as long as the side surfaces are gapped (see further discussion on this in Sec.~\ref{sec:3c}), there should be HOMAs in the rod Hamiltonian $H_{\text{BdG}}(x,y,k_z)$ in these $k_z$ regions. However, here due to an effective 2D particle-hole symmetry $\mathcal{\tilde P'}\equiv \mathcal{\tilde PM}_z $ of the BdG Hamiltonian, the HOMAs are pinned exactly at zero energy.

\begin{figure}
\subfloat[]{\includegraphics[width=0.8\columnwidth]{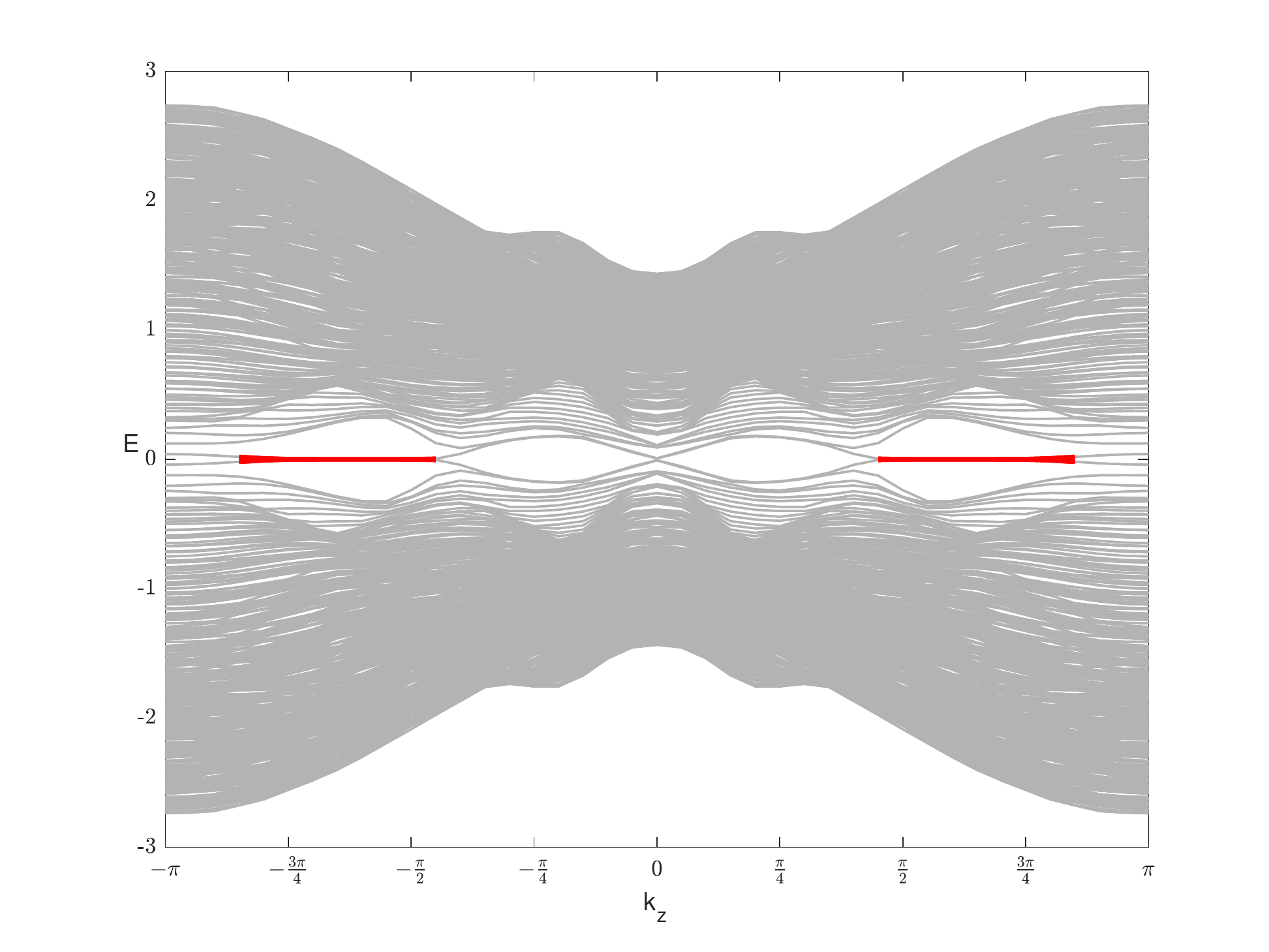}}
\hfill
\subfloat[]{\includegraphics[width=0.8\columnwidth]{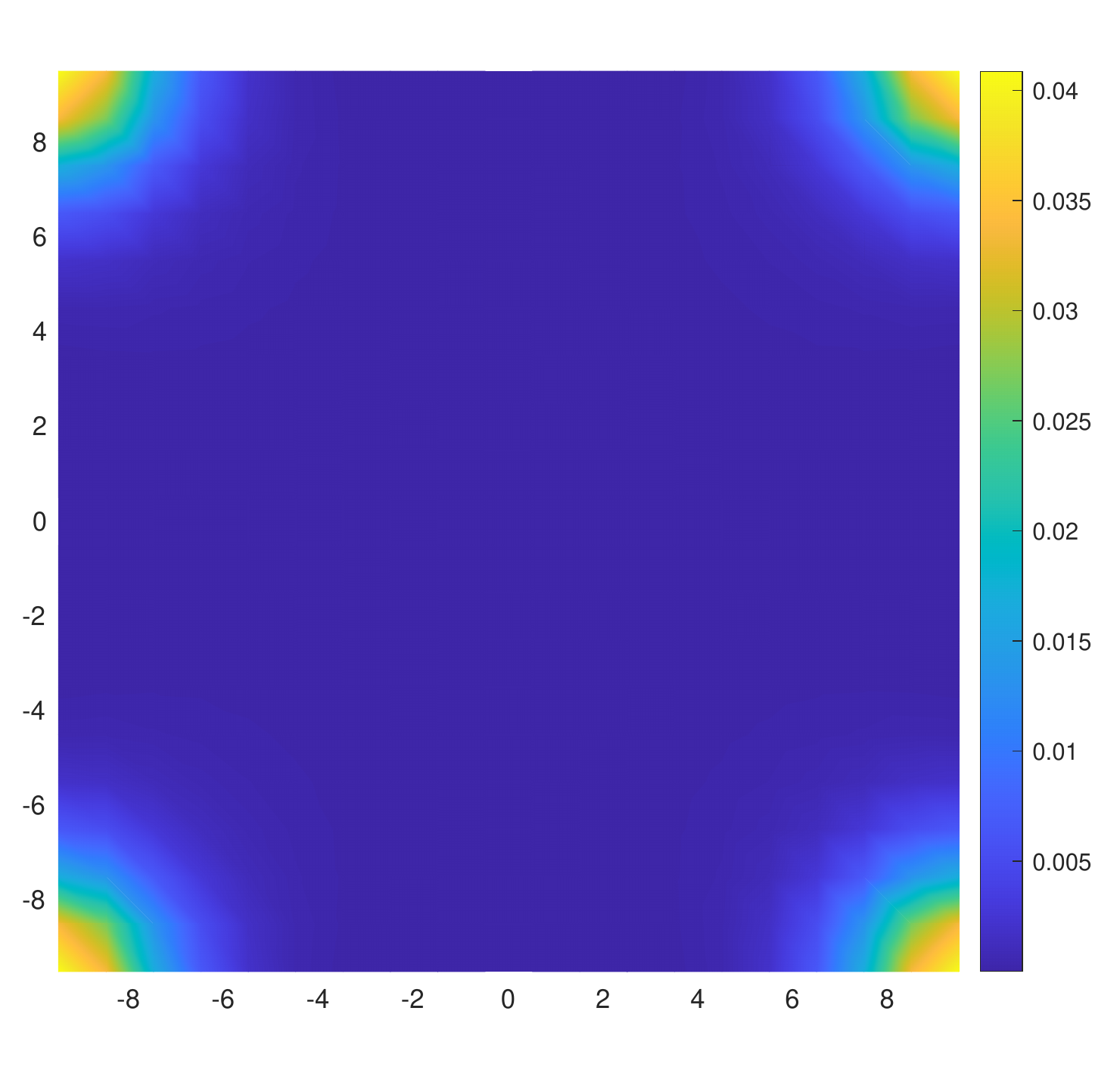}}
\caption{(a) Energy spectrum of the BdG rod Hamiltonian $H_{\text{BdG}}(x,y,k_z)$ as a function of $k_z$ for $N = 20$. Red lines represent the mid-gap Majorana zero modes on hinges. (b)Wave-function distributions of the Majorana corner states on a 2D slice in (a) at $k_z = 2\pi/3$. Parameters are $M_0 = 1.5$, $t_{xy} = t_z = \eta = \lambda_1 = 1$, $\lambda_2 = 0$ and $\mu = \Delta_1  = 0.4$.}
\label{fig:6}
\end{figure}

\begin{figure}
    \includegraphics[width = 2in]{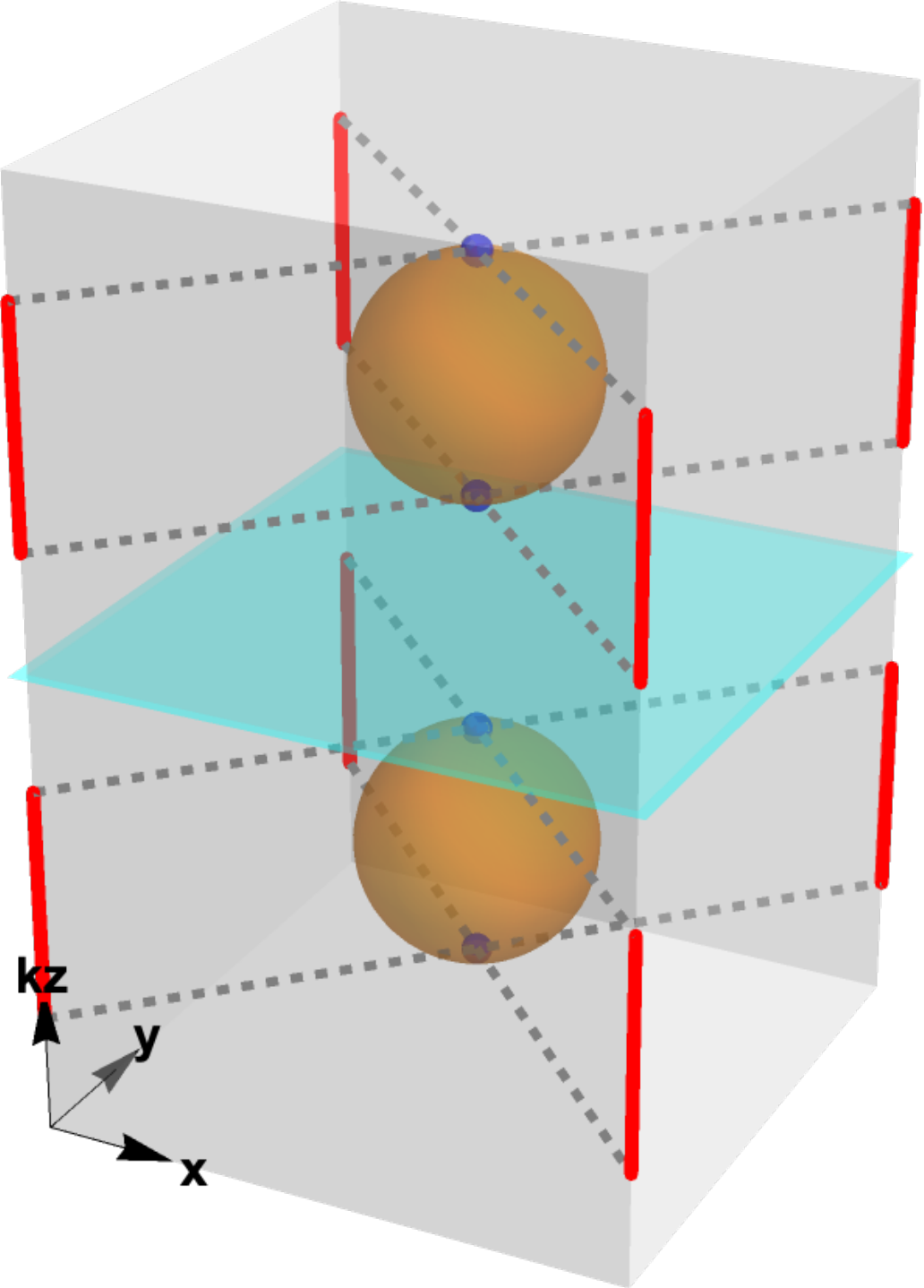}
    \caption{A schematic plot of the $\mathcal{C}_4$ symmetric Dirac superconductor phase with hybrid  higher-order topology. Two brown spheres represent a pair of disjointed FSs and four blue dots denote the bulk BdG Dirac points on the North poles and South poles of two FSs. Red lines denote HOMAs which sink into the bulk at hinge projections of the four Dirac points. The cyan plane at $k_z = 0$ carries a nontrivial mirror Chern number, ensuring surface Majorana zero modes.}
    \label{fig:7}
\end{figure}

We numerically solve the BdG Hamiltonian \eqref{eq:17} and the results that visualize the corner Majorana modes are shown in Fig. \ref{fig:6}. At $k_z = 0$, the gapless surface modes are due to the mirror Chern number, as we mentioned [see Fig. \hyperref[fig:6]{\ref{fig:6}(a)}].
 The 3D rod Hamiltonian $H_{\text{BdG}}(x,y,k_z)$ displays hinge HOMAs that terminate at the projections of bulk BdG Dirac points, together with surface Majorana helical states at the nontrivial mirror plane $k_z = 0$. Generally, the HOMAs in this $\mathcal{C}_4$-symmetric Dirac superconductor are disconnected (see Fig. \ref{fig:7}).

As a comparison, a higher-order topological Dirac superconductor phase protected by $\mathcal{C}_6$ rotation and inversion symmetry is introduced in Ref.\ \cite{HOTDSC}, which hosts 3D bulk Dirac nodes, 2D gapped surface states and 1D HOMAs simultaneously in the BdG spectrum. The authors also propose an inversion symmetry indicator $\kappa_{2d}$ at $k_z = 0$ to characterize the 1D HOMAs. However, this symmetry indicator as a justification of HOMAs does not apply to our case because HOMAs do not cross the $k_z = 0$ plane in general. The $\mathbb{Z}_2$ higher-order topological invariant on a $k_z$ slice Hamiltonian defined in Eq. (\ref{eq:28}) is sufficient to diagnose the HOMAs, as long as the surface is gapped. Our $\mathcal{C}_4$-symmetric Dirac superconductor instead displays a hybrid higher-order topology \cite{HOTSC_hybrid}.

\subsection{Irremovability of HOMAs due to additional $\mathbb{Z}$-valued monopole charges}
\label{sec:3d}

In Sec.~\ref{sec:3b} we have identified a nontrivial $\mathbb{Z}_2$ monopole charge hosted by each of the four BdG Dirac points. If that were the only the monopole charge hosted by a BdG Dirac point, then \emph{any} of the two Dirac points could annihilate each other. In particular, if the BdG Dirac points at $(0,0,k_1)$ and $(0,0,k_2)$ annihilated, the HOMA terminating at $k_z=k_1$ and $k_2$ would disappear.  Since for small Fermi surfaces (small $\mu$) $k_1$ and $k_2$ are close, this naturally raises the question whether the BdG Dirac points and HOMAs are stable beyond small perturbations.

Fortunately, the BdG Dirac points are stabilized by an additional monopole charge $\in \mathbb{Z}$ protected by the $\mathcal{C}_4$ symmetry. To this end, we note that for each $k_z$ layer
\begin{align}
\lambda (k_z) = N_{J =1/2}(0,0,k_z) \in \mathbb{Z},
\label{eq:lambda}
\end{align}
i.e., the occupation number of the band with $J =1/2$ at the $\Gamma$ point, is a topological invariant which cannot  change without a gap closing. From Fig.~\hyperref[fig:5]{\ref{fig:5}(a)} it is straightforward to read off that for $\lambda=0$ for  $|k_z|< k_1$, $\lambda=1$ for $k_1<|k_z|<k_2$, and $\lambda=2$ for $|k_z|>k_2$.

We can thus similarly assign a monopole charge $\Delta\lambda =1 $ for the BdG Dirac points at $k_z= k_{1,2}$, and $\Delta\lambda =-1$ for $k_z= -k_{1,2}$ to the BdG Dirac points. Importantly, we see that the BdG Dirac points originating from the same normal state Dirac point, e.g., with $k_z = k_1$ and $k_2$, cannot annihilate, since they carry a total $\mathbb{Z}$ monopole charge $\Delta\lambda =2$~\footnote{The BdG Dirac points can only annihilate with those from another Dirac point. However, that does not happen in the $\Delta\ll \mu$ limit for the Dirac semimetal normal state.}.  Just like the $\mathbb{Z}_2=1$ monopole charge $\Delta\nu$ we showed in Sec.~\ref{sec:3b},  the origin of $\Delta\lambda=2$ can be traced back to the nonzero angular momentum carried by the $B_{1u}$ and $B_{2u}$ pairing order.

Thus we conclude that the BdG Dirac points are stable due to the additional $\mathbb{Z}$ monopole charge protected by $\mathcal{C}_4$ symmetry. As a result, the HOMA connecting the BdG Dirac points is a stable spectral feature that cannot be removed by perturbations. Indeed, in the BdG Hamiltonian Eq.~\eqref{eq:17}, as $\mu\to 0$ the two BdG Dirac points with the same sign of $k_z$ coincide, but as $\mu$ changes sign the two BdG Dirac points simply cross each other without annihilating each other.

In spirit, our argument here is similar to the doubly-charged topological nodes studied in Ref.~\cite{NLSM3} with inversion symmetry only, where an additional topological monopole charge protects the nodal structure from annihilating itself.  Our argument goes beyond the classification there, since the $\mathcal{C}_4$ rotation symmetry plays an important role in defining the additional monopole charge and protecting the HOMAs. \\

\subsection{Mirror-protected gapless surface states and HOMAs }
\label{sec:3c}

While the higher-order topology and BdG Dirac points are protected by $\mathcal{C}_4$ and $\mathcal{TI}$ symmetries alone, the system has many additional spatial symmetries, such as mirror symmetries. As the existence of HOMAs requires the surface states to be gapped, the surfaces need to avoid the orientations in which there are mirror-protected gapless surface states. In this subsection, we show that the $B_{1u}$ state exhibits robust gapless states on the side surfaces in the $(x\pm y),z$ (diagonal) direction, and similarly for the $B_{2u}$ state on surfaces in the $xz$ and $yz$ directions. Therefore, $B_{1u}$ and $B_{2u}$ states exhibit HOMAs in different geometrical configurations. Moreover, we show that the HOMAs obtained previously have a close relationship to these surface states from the perspective of mirror symmetries.

Let us first focus on the $B_{2u}$ state. For definiteness, we consider a 2D subsystem with $k_z=k_0$, a plane that contains the normal-state Dirac point and is topologically nontrivial. Without loss of generality, we set $\mu=0$, and from Eq.~\eqref{eq:17}, the BdG Hamiltonian can be written  as
\begin{align}
H_{\rm BdG, 2D}(k_x, k_y) =& (2-\cos k_x - \cos k_y) \sigma_z\tau_z + \Delta_2 \sigma_ys_z \tau_x\nonumber\\
&+ \sin k_x \sigma_xs_z - \sin k_y \sigma_y\tau_z  \nonumber\\
 &+ \lambda_1'(\cos k_y - \cos k_x)\sigma_xs_x \nonumber\\
&+ \lambda_2' \sin k_x\sin k_y \sigma_x s_y,
\end{align}
where $M_0' = 2$ at $k_z = k_0$. We consider the edge of the system along the $x$ and $y$ directions. For the {$x$} edge, $k_y$ is a good quantum number and we set $k_y=0$. The 1D Hamiltonian is
\begin{align}
H_{\rm 1D}(k_x) =& (1-\cos k_x) \sigma_z\tau_z + \Delta_2 \sigma_ys_z \tau_x\nonumber\\
&+ \sin k_x \sigma_xs_z + \lambda_1'(1 - \cos k_x)\sigma_xs_x,
 \label{eq:BDI}
\end{align}
which is symmetric under several composite symmetries, including an effective time-reversal symmetry 
\begin{align}
 \mathcal{\tilde T}' =  \mathcal{\tilde T}\mathcal{\tilde M}_x\mathcal{\tilde M}_y \mathcal{\tilde I} = s_x \mathcal{K}
 \end{align}
 and an effective particle-hole symmetry 
 \begin{align}
 \mathcal{\tilde P}' =  \mathcal{\tilde P}\mathcal{\tilde M}_x\mathcal{\tilde I} = \sigma_z s_x \tau_x \mathcal{K}
 \end{align}
satisfying $ \mathcal{\tilde T}'^2 = +1$  and $ \mathcal{\tilde P}'^2 = +1 $. The two symmetries together place the 1D subsystem $H_{\rm 1D}(k_x) $ in class BDI~\cite{qin_zhang_wang_fang_zhang_hu,Tewari_Sau_BDI,NC_BDI,AIP_Ryu,Ryu_IOP,Ryu_PRB}, which admits a $\mathbb{Z}$ classification given by a winding number. Via a direct computation (see Appendix \ref{appen:2}), we find that the winding number is $w=2$, corresponding to two Majorana zero modes. Extended to finite $k_y$, this indicates that the {$x$} edge hosts helical Majorana states. Therefore, the $yz$ and $xz$ (related to $yz$ by a $\mathcal{C}_4$ rotation) surfaces for the $B_{2u}$ state are gapless.
  
  Similar considerations show that the $B_{1u}$ state also hosts gapless surfaces. While $B_{1u}$ and $B_{2u}$ order parameters transform differently under mirror reflection, it can be directly verified that the $(x\pm y), z$ surfaces are gapless, protected by the diagonal reflection symmetries $\tilde{\mathcal{M}}_{x\pm y}\equiv \tilde{\mathcal{M}}_{x, y}\mathcal{C}_4$.
  
As the existence of HOMAs requires gapped surfaces, we conclude that the $B_{1u}$ state hosts  HOMAs in a ``square cut" (with $xz$ and $yz$ surfaces), and the $B_{2u}$ state hosts HOMAs in a ``diamond cut" [with $(x\pm y),z$ surfaces]. This is the central result of this work, illustrated in Table~\ref{table:0}.
  
We note that the existence of HOMAs can be analytically understood from the gapless surfaces as well, using an argument similar to that in Sec.~\ref{sec:2b}. Namely, one can view the hinge as a curved surface in the extreme limit. For the $B_{2u}$ state, as the surface curls away from the $yz$ direction, the surface helical states acquire a mass, and by continuity the induced mass should change sign as the curved surfaces turn from $(x+y),z$ direction to $(x-y),z$ direction. Therefore, at the mass domain wall there exists a Majorana zero mode, which is localized at the hinge.

\begin{table*}
\begin{tabular}{c|c|c|c|c|c|c}
    \hline
    \hline
    Representative pairing order& Irrep & $\mathcal{C}_4 = is_z e^{i\frac{\pi}{4}\sigma_z s_z}$ & $\mathcal{I} = \sigma_z$ & $\mathcal{M}_x = is_x$ & $\mathcal{M}_y = i\sigma_z s_y$ & $\mathcal{M}_z = is_z$\\
    \hline
    $\Delta_1 \sigma_y$  & $B_{1u}$  & $-$    & $-$ &   $-$ & $-$ & $-$\\
    $\Delta_2 \sigma_y s_z$& $B_{2u}$  &   $-$  & $-$  & $+$ & $+$ & $-$\\
    $\Delta_3 k_z s_x, \Delta_4 k_z \sigma_z s_x$& $A_{1u}$  &   $+$  & $-$  & $-$ & $-$ & $-$\\
    $\Delta_5 k_x k_y\sigma_y, \Delta_6 (k_x^2-k_y^2)\sigma_y s_z$& $A_{2u}$  &   $+$  & $-$  & $+$ & $+$ & $-$\\
    $\Delta_7 (\sigma_y s_x,\sigma_x s_y),\Delta_8 k_z (\sigma_z , s_z)$& $E_{u}$  &   $0$  & $-$  & $0$ & $0$ & $+$\\
    $\Delta_9k_z\sigma_x s_z$& $B_{1g}$  &   $-$  & $+$  & $+$ & $+$ & $+$\\
    $\Delta_{10}k_z\sigma_x$& $B_{2g}$  &   $-$  & $+$  & $-$ & $-$ & $+$\\
    $\Delta_{11}s_y, \Delta_{12}\sigma_z s_y$& $A_{1g}$  &   $+$  & $+$  & $+$ & $+$ & $+$\\
    $\Delta_{13}k_z(k_x^2-k_y^2)\sigma_x, \Delta_{14}k_z k_x k_y\sigma_x s_z$& $A_{2g}$  &   $+$  & $+$  & $-$ & $-$ & $+$\\
    $\Delta_{15}k_z(\sigma_x s_x, \sigma_y s_y)$& $E_{g}$  &   $0$  & $+$  & $0$ & $0$ & $-$\\
    \hline
    \hline
   \end{tabular}
\caption{List of representative pairing orders {(to the lowest order in $\bf k$)} {for all  irreducible representations of $D_{4h}$}, and their corresponding characters. Entries with 0 correspond to 2D representations with opposite eigenvalues. }
\label{table:2}
\end{table*}

\subsection{Symmetry breaking effects}
\label{sec:3e}

Like those in the normal state, the BdG Dirac points and the HOMAs are stabilized by $\mathcal{TI}$ and $\mathcal{C}_4$ symmetries. The mirror symmetries we discussed in Sec.~\ref{sec:3c} can be broken by adding a small $\sin k_x\sin k_y \sin k_z$ component to the $\lambda_1$ term, or by mixing $B_{1u}$ and $B_{2u}$ states, and gapless surface modes will be removed. Nevertheless, the BdG Dirac points and the HOMAs remain robust due to the remaining $\mathcal{C}_4$ symmetry.

In a system with an additional inversion-breaking term such as  $\sin k_z s_z$, or an additional time-reversal-breaking Zeeman term~\cite{monopoleDSM} such as $m s_z \tau_z$, the BdG Dirac points spilt into two Weyl points with opposite chiralities along the $z$ axis. Surface Majorana arcs appear in $k_z$ regions between Weyl points with opposite chiralities because the corresponding 2D subsystem carries a nontrivial Chern number and it is not Wannier representable. Since HOMAs are locally protected by a 2D version of particle-hole symmetry $\mathcal{P}\mathcal{M}_z$, the hinge modes are stable under time-reversal-breaking perturbations. Therefore, the $\mathcal{T}$-breaking phase is a higher-order Weyl superconductor defined in Ref. \cite{HOTDSC}, in which HOMAs and surface Majorana arcs coexist. On the other hand, an inversion-breaking term breaks $\mathcal{M}_z$, and correspondingly destroys HOMAs.

Moreover,  $\mathcal{C}_4$ symmetry can be broken down to $\mathcal{C}_2$, e.g., by mixing a $A_{1u}$ pairing term $\sin k_z s_x\tau_x$ with $B_{1u}$, which gaps out the bulk BdG Dirac points. The 2D slice Hamiltonian thus becomes a boundary obstructed topological superconductor~\cite{TiwariJahinWang}, in which corner Majorana modes still exist and are protected by the surface gap instead of the bulk gap. In this case the bulk is fully gapped and the HOMA terminates at a surface gap-closing point instead.

\section{Topological properties of other pairing states}
\label{sec:4}

\begin{figure}
	\subfloat[]{\includegraphics[width=0.4\columnwidth]{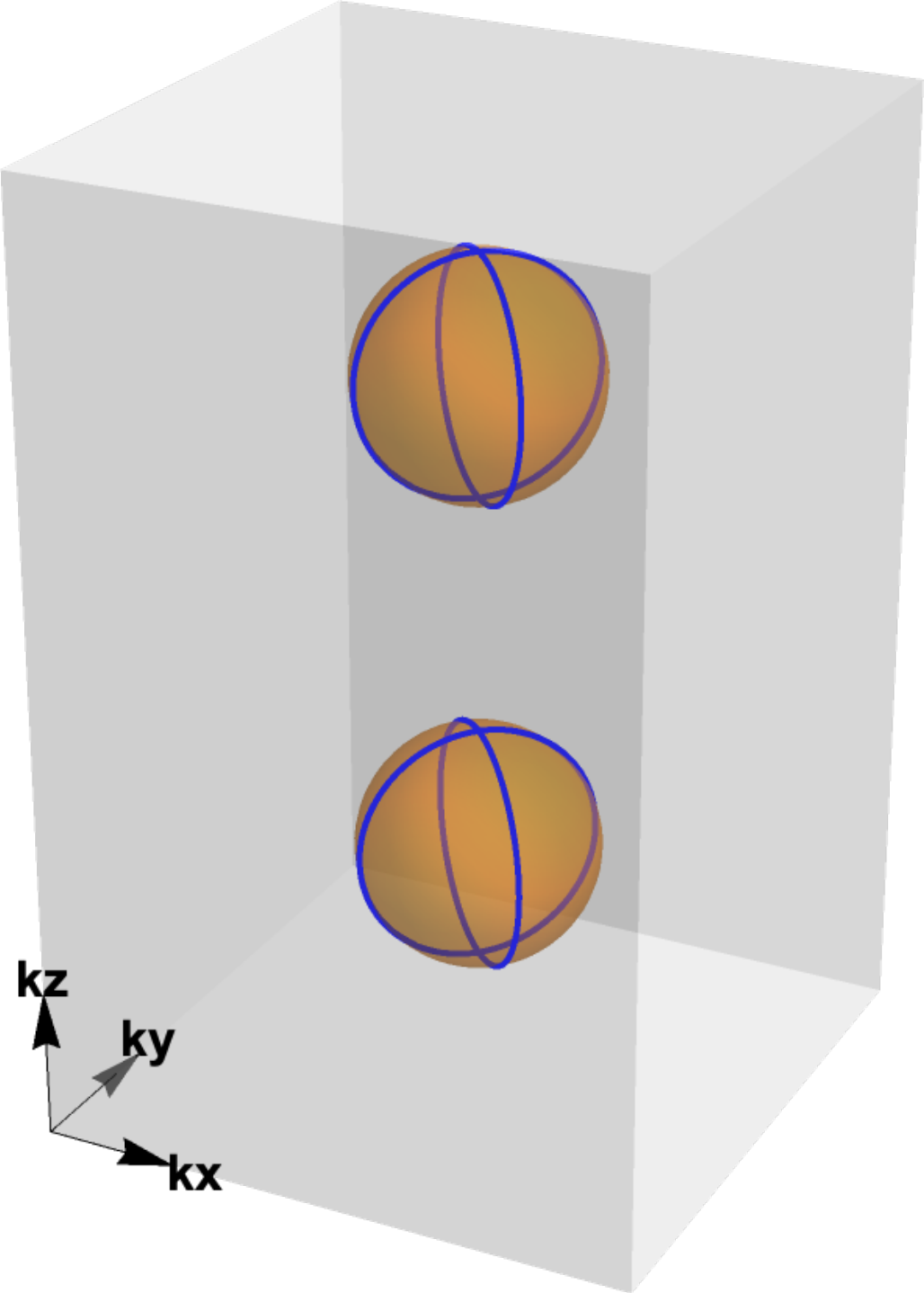}}
	\hfill
	\subfloat[]{\includegraphics[width=0.4\columnwidth]{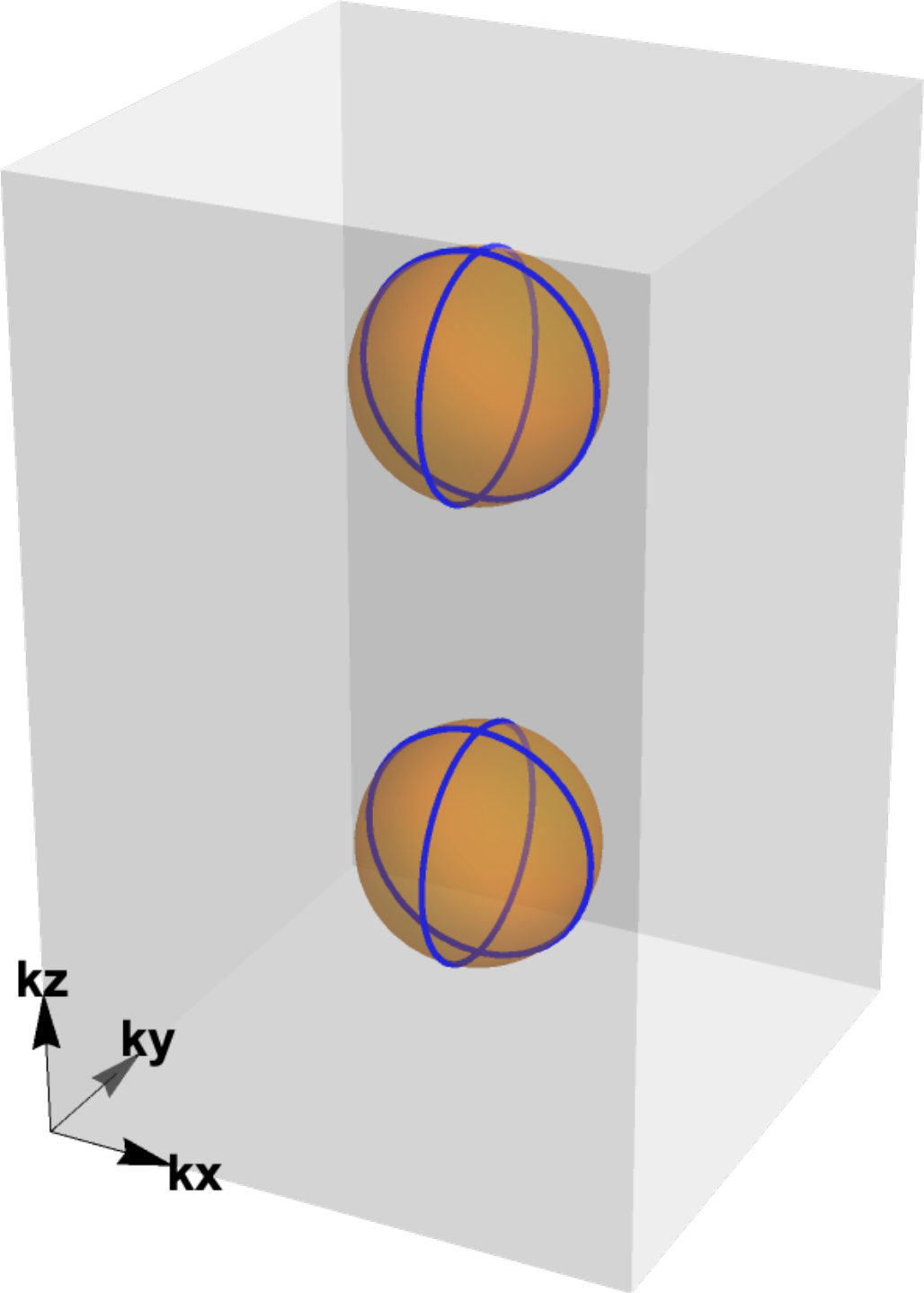}}
   \caption{Schematic plots of bulk nodal cages(marked blue) in the 3D BZ for (a) $B_{1g}$ and (b) $B_{2g}$ channels, both $\mathcal{C}_4$ symmetric. Brown spheres denote Fermi surfaces. }
\label{fig:Bg}
\end{figure}

So far we have focused on the $B_{1u}$ and $B_{2u}$ pairing channels in a Dirac semimetal. Remarkably, we have shown the topological properties are largely determined by their symmetry properties, in particular, the transformation of the pairing order under the $\mathcal{C}_4$ rotations. In this section we briefly discuss the topological properties of the system in other pairing channels.

In Table~\ref{table:2} we list all irreducible representations of the point group $D_{4h}$ and their representative order parameters to the lowest orders in $\bf k$. {However, as is the case for $B_{1u}$ and $B_{2u}$, the analysis of their topological properties does not depend on the detailed $k$ dependence of the order parameters but rather only on their symmetry properties, unless there are accidental nodes.}

We note that $B_{1g}$ and $B_{2g}$ orders transforms under $\mathcal{C}_4$ the same way as $B_{1u}$ and $B_{2u}$, but are of even parity. Direct diagonalization of the corresponding BdG Hamiltonians shows that the system exhibits $\mathcal{C}_4$-symmetric cages of nodal lines~\cite{Nodalcage} in both pairing states between $k_1<|k_z|<k_2$, shown in Fig.~\ref{fig:Bg}. These nodal lines are four-fold degenerate, and are protected by a $2\mathbb{Z}$ topological invariant in the AZ+$\mathcal{I}$ classification table~\cite{NLSM3}. The gaplessness of the 2D subsystem for $k_1<|k_z|<k_2$ renders the higher-order topological invariant $\nu$, and consequently the HOMAs, ill defined. However, the  monopole charge $\Delta\lambda = \lambda(k>k_2)-\lambda(k<k_1)$, is still valid. As $\lambda$ in Eq.~\eqref{eq:lambda} is solely determined by $\mathcal{C}_{4}$ operations, which is identical for $B_{1g}/B_{2g}$ and for $B_{1u}/B_{2u}$, we conclude, by a similar analysis to that in Secs.~\ref{sec:3b} and \ref{sec:3d}, that the nodal cage also carries a monopole charge of $\Delta\lambda =2$. For this reason, the nodal cages are stable and cannot self-annihilate. Indeed, as $\mu$ goes through zero, the nodal cages shrink to {a point} but then reemerges. In this case they are protected by a doubly charged monopole beyond the classification in Ref.~\cite{NLSM3}.

The entries ($E_u$ and $E_g$) with characters $0$ indicates these irreps are two dimensional. Indeed, they have angular momenta $\pm 1$, each of which breaks time-reversal symmetry. In these cases the system either breaks time-reversal or $\mathcal{C}_4$ symmetry, but cannot preserve both. Yet, they may still exhibit interesting topological properties. However, we will restrict ourselves to time-reversal  and $\mathcal{C}_4$-invariant systems in this work and leave a detailed analysis of the $E_u$ and $E_g$ states to a future study.

All other pairing channels transform trivially under $\mathcal{C}_{4}$ rotation, and it is straightforward to show that they do not carry any nontrivial $\mathbb{Z}_2$ or $\mathbb{Z}$ monopole charges in relation to HOMAs studied here. A recent study \cite{Yan_HOTSC} investigated a $A_{1g}$ pairing channel  with a special form factor in doped Dirac semimetals. The authors found a second-order topological superconducting phase with helical Majorana modes on the top and bottom hinges. However, this nontrivial phase requires a fine tuning of model parameters.
The conclusion that a Dirac semimetal is topologically obstructed only for certain pairing channels was also obtained in Ref.~\cite{Sun_Z2}. However, the symmetries in consideration are different from those in our case.

Finally, while we demonstrated from a microscopic calculation that $B_{1u}$ and $B_{2u}$ pairing orders are naturally induced by Ising-ferromagnetic fluctuations, an interesting open problem is what microscopic interactions promote other nontrivial pairing channels, such as $B_{1g}$ and $B_{2g}$.

\section{Summary}
\label{sec:5}
In this work, we  analyzed the higher-order topology in superconducting Dirac semimetals protected by fourfold-rotation, spatial-inversion and time-reversal symmetries. In the normal state, the Dirac semimetal is known~\cite{NC_HOFA,Fang_classifyHOFA,Fang_fillinganomaly} to exhibit HOFAs. We showed that in certain pairing channels the normal state topology gets inherited in the superconducting state.

First, we showed that in the presence of an Ising-ferromagnetic fluctuations, the leading pairing instability of the system is toward $B_{1u}$ or $B_{2u}$ pairing orders. These pairing states display Dirac nodal points in the BdG spectrum. In a rod geometry, they display HOMA states at the hinges that terminate at the projections of bulk BdG Dirac points onto the hinges. While HOFAs are generally dispersive in the energy spectrum, HOMAs are pinned at zero energy due to the particle-hole symmetry. Importantly, the BdG Dirac points and the HOMAs are protected by an additional $\mathbb{Z}$-valued monopole charge defined via the $\mathcal{C}_4$ symmetry.

Moreover, the recipe for diagnosis of higher-order topology and monopole charge here can be directly applied to all other pairing orders, which are irreps of the $D_{4h}$ group. In particular, we found that $B_{1g}$ or $B_{2g}$ displays nodal cages with a monopole charge, each of which is stable under $\mathcal{C}_4$-symmetric perturbations.

In this work we showed that Dirac semimetals are promising ``parent states" for higher-order topological superconducting states. {Since HOFAs have been experimentally confirmed in the Dirac semimetal Cd$_3$As$_2$~\cite{HOFAexp1,HOFAexp2},  it would be interesting to search for nontrivial pairing states in these systems.

\acknowledgements
We thank Y. Li , S. Qin, J. Zhang and A. Jahin for illuminating discussions. This work is supported by startup funds at the University of Florida and by NSF under Award No. DMR-2045781.\\

\appendix

\section{{Wyckoff positions from $\mathcal{C}_4$ eigenvalues at $\Gamma$ and $M$}}
\label{appen:0}
\begin{figure}
	\subfloat[$\Gamma$]{\includegraphics[width=0.8\columnwidth]{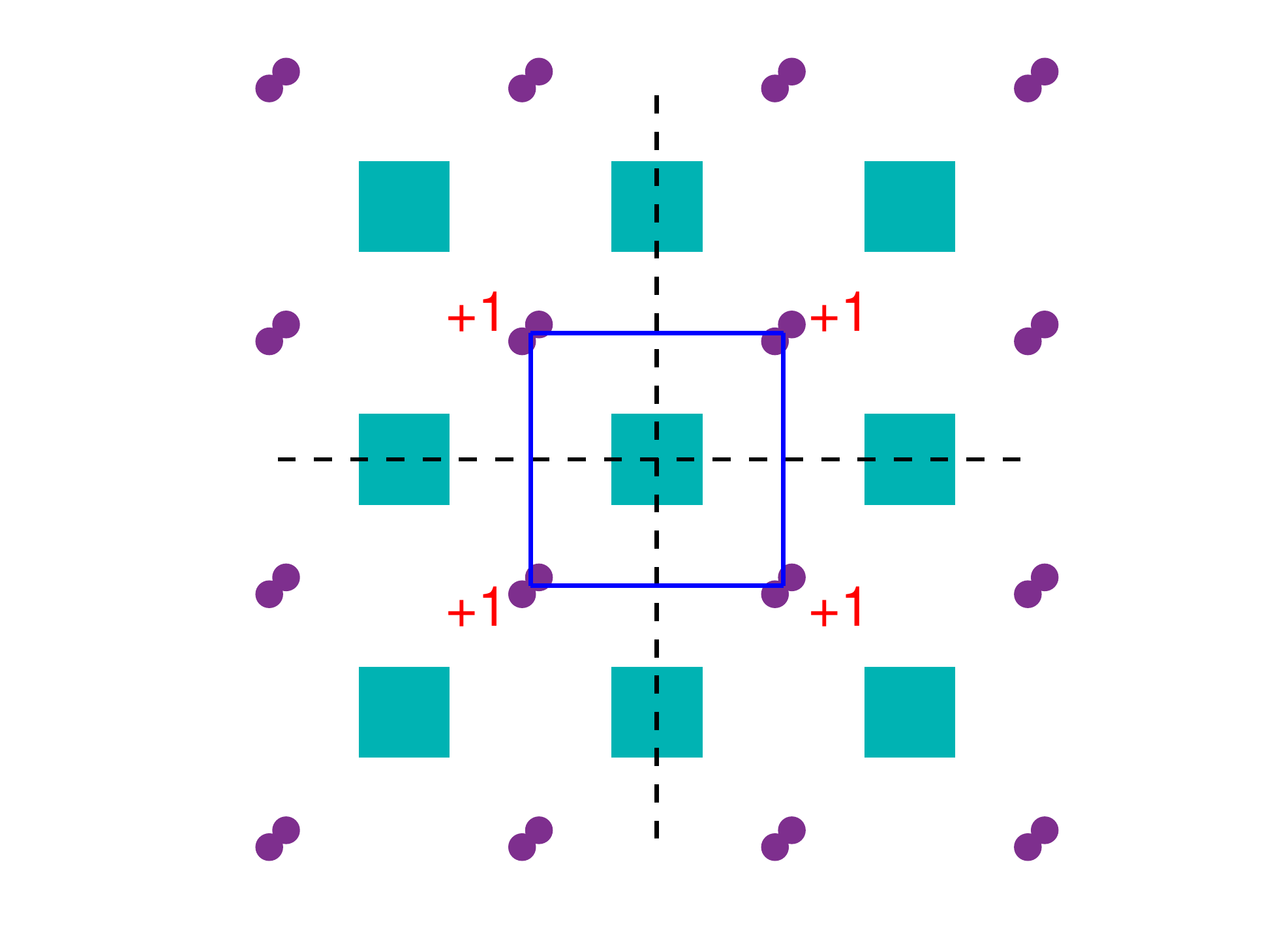}}
	\hfill
	\subfloat[$M$]{\includegraphics[width=0.8\columnwidth]{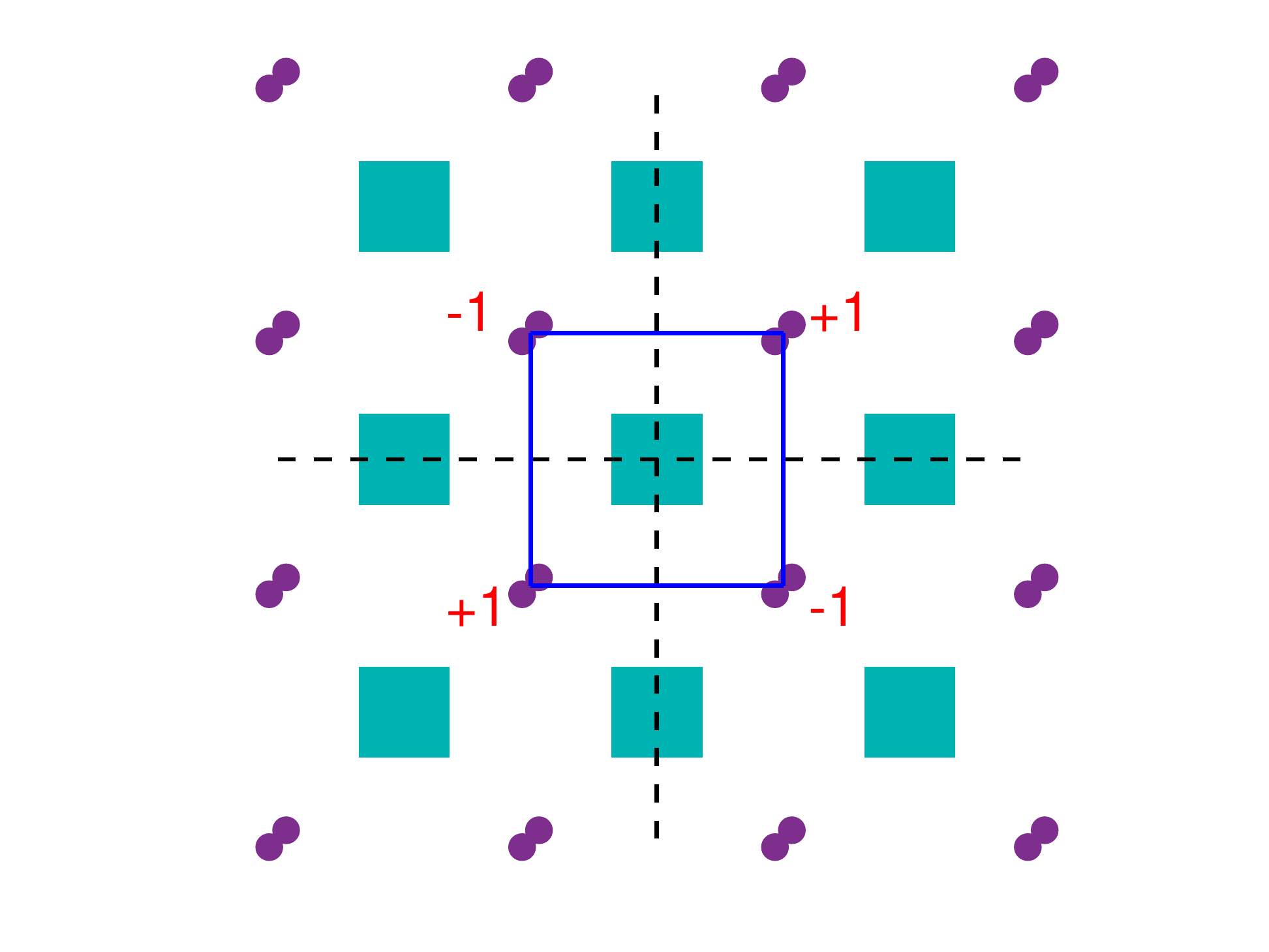}}
   \caption{The relative phases (marked red) of spatial parts of Wannier functions centered at Wyckoff positions ${\bf r} = (\pm\frac12,\pm\frac12)$ at (a) $\Gamma$ and (b) $M$. The blue square is a unit cell and we assume no edges.}
\label{fig:Wyckoff1b}
\end{figure}

{In this appendix, we briefly explain how to get Fig. \ref{fig:normal} from $\mathcal{C}_4$ eigenvalues at high-symmetry points $\Gamma$ and $M$ in BZ. For Wannier functions centered at the Wyckoff position ${\bf r} = (\frac12,\frac12)$, we plot the relative phases of spatial parts of Wannier functions at four $\mathcal{C}_4$ related corners of the unit cell in Fig. \ref{fig:Wyckoff1b}, for $\Gamma$ and $M$ separately. Under $\mathcal{C}_4$ rotation, the spatial part of the Wannier function at $M$ acquires a minus sign compared with $\Gamma$.}

{From Eq. \eqref{eq:6}, we obtain that the nontrivial phase [$\nu = 1$ defined in Eq. \eqref{eq:8}] corresponds to $\mathcal{C}_4 = e^{\pm i\frac{\pi}{4}}$ for the two occupied bands at $\Gamma$ and $\mathcal{C}_4 = e^{\mp i\frac{3\pi}{4}}$ for the two occupied bands at $M$. With a relative minus sign for $\mathcal{C}_4$ eigenvalues at $\Gamma$ and $M$, the real-space picture matches the configuration plotted in Fig. \ref{fig:Wyckoff1b}. Note that the exact eigenvalue of $\mathcal{C}_4$ comes from both the spatial part and the intrinsic part (e.g., orbital or spin indices) of the Wannier function and the latter is not important when determining the Wannier centers.}

{On the other hand, the trivial phase ($\nu$ = 0) corresponds to the same $\mathcal{C}_4$ eigenvalues at $\Gamma$ and $M$ so that the Wannier functions are centered at ionic sites ${\bf r} = (0,0)$ which is an atomic limit without filling anomaly.}

\section{Derivations of form factors $F_{1,2}({\bf k})$}
\label{appen:1}
We simplify the form of the normal state Hamiltonian $H_0({\bf k})$ in  Eq. \eqref{eq:1} as
\begin{align}
 H_0({\bf k}) =&  f_1({\bf k})\sigma_z + f_2({\bf k})\sigma_xs_z + f_3({\bf k})\sigma_y\nonumber\\
 &+ f_4({\bf k})\sigma_xs_x + f_5({\bf k})\sigma_xs_y.
 \label{eq:A1}
\end{align}
The Green's function for finite $\mu$ is given by 
\begin{align}
G_0 (i\omega_m,{\bf k}) &= [i\omega_m - H_0({\bf k}) + \mu]^{-1}\nonumber\\
&= \frac{(i\omega_m+\mu)\sigma_0s_0 +H_0({\bf k})}{(i\omega_m+\mu)^2 - \epsilon^2},
\end{align}
where $\omega_m = (2m+1)\pi T$ is the fermionic  Matsubara frequency and  $\epsilon = \sum_{i=1}^{5}\sqrt{f_{i}^2({\bf k})}$ is the dispersion. The trace in Eq. \eqref{eq:gap1} can be explicitly evaluated as (e.g., for $\Delta_1$)
%\begin{widetext}
\begin{align}
&\text{Tr}\[ \sigma_y G_0 (i\omega_m,{\bf k}) \sigma_y G_0^{\text{T}} (-i\omega_m,-{\bf k})\] \nonumber\\
=& \text{Tr}\left\{[(i\omega_m+\mu)\sigma_0s_0+\sigma_y H_0({\bf k})\sigma_y]\right. \nonumber\\
&\times\left.[(-i\omega_m+\mu)\sigma_0s_0+H_0^{\text{T}}(-{\bf k})]\right\} \nonumber\\
&\times\frac{1}{[(i\omega_m+\mu)^2-\epsilon^2][(-i\omega_m+\mu)^2-\epsilon^2]}\nonumber\\
 =& \frac{\text{Tr}\left\{(\omega_m^2 + \mu^2)\sigma_0s_0+\sigma_yH_0({\bf k})\sigma_yH_0^{\text{T}}(-{\bf k})\right\}}
{[\omega_m^2 + (\epsilon + \mu)^2][\omega_m^2 + (\epsilon-\mu)^2]}\nonumber\\
 =& \frac{4[\omega_m^2+\mu^2-\epsilon^2 + 2(f_2^2({\bf k}) + f_3^2({\bf k}) + f_4^2({\bf k}))]}
{[\omega_m^2 + (\epsilon + \mu)^2][\omega_m^2 + (\epsilon-\mu)^2]}\nonumber\\
 \simeq &\frac{2[f_2^2({\bf k}) + f_3^2({\bf k}) + f_4^2({\bf k})]}
{\mu^2[\omega_m^2 + (\epsilon-\mu)^2]},
\end{align}
%\end{widetext}
where we take $\epsilon \simeq \mu$ in the last step. The sum over Matsubara frequencies yields 
\begin{align}
&T_{c1}\sum_{\omega_m}\frac{2[f_2^2({\bf k}) + f_3^2({\bf k}) + f_4^2({\bf k})]}
{\mu^2[\omega_m^2 + (\epsilon-\mu)^2]} \nonumber\\
=& \frac{f_2^2({\bf k})+f_3^2({\bf k})+f_4^2({\bf k})}{\mu^2}\frac{\tanh\frac{\epsilon-\mu}{2T_{c1}}}{\epsilon-\mu}.
\end{align}
There are two nearly spherical Fermi pockets surrounding the Dirac points at $(0,0,\pm k_0)$. By expanding $f_2({\bf k})$, $f_3({\bf k})$ and $f_4({\bf k})$ in lowest order of momentum, we obtain the corresponding $F_1({\bf k})$ in the main text. $F_2({\bf k})$ can be derived similarly.

\section{Topological invariant of the 1D Hamiltonian $H_{\rm 1D}(k_x)$}
\label{appen:2}
We follow Ref. \cite{Tewari_Sau_BDI} to derive the topological invariant of the Hamiltonian in Eq. \eqref{eq:BDI}. Note that $H_{\rm 1D}(k_x)$ has a chiral symmetry defined as $\mathcal{\tilde S} =  \mathcal{\tilde T}' \mathcal{\tilde P}' = \sigma_z\tau_x$, such that $\mathcal{\tilde S}H_{\rm 1D}(k_x)\mathcal{\tilde S}^{-1} = -H_{\rm 1D}(k_x)$. It can be made purely off-diagonal by a unitary transformation in the particle-hole space 
\begin{align}
\tilde{H}(k_x) = UH_{\rm 1D}(k_x)U^{\dag} = \[
\begin{matrix}
 & A(k_x)\\
A^{\dag}(k_x) & 
\end{matrix}
\],
\end{align}
where $U = e^{-i\frac{\pi}{4}\tau_y}$ and we obtain 
\begin{align}
A(k_x) =& (1-\cos k_x)s_0\nu_z + (\sin k_x + i\Delta_2) s_z \nu_x  \nonumber\\
&+ \lambda_1'(1-\cos k_x)s_x\nu_x,
\end{align}
where $\nu_{x,y,z}$ are Pauli matrices spanning the subspace $\langle\sigma_z\tau_z = 1|\sigma_i\tau_j|\sigma_z\tau_z=-1\rangle$. The topological invariant in BDI class is defined as the winding number of the phase factor carried by ${\rm det}A(k_x)$, which is 
\begin{align}
w = \frac{-i}{\pi}\int_{k_x = 0}^{k_x = \pi} \frac{dz(k_x)}{z(k_x)},
\end{align}
where $z(k_x) = e^{i\theta(k_x)} = {\rm det}A(k_x)/|{\rm det}A(k_x)|$. Note that 
\begin{align}
&{\rm det}A(k_x)\\
=& \[
(1+\lambda_1'^2)(1-\cos k_x)^2 +\sin k_x^2 -\Delta_2^2 + 2i\Delta_2\sin k_x
\]^2 \nonumber
\end{align}
and $\Delta_2\ll 1$ in the weak pairing limit. Hence $\theta(k_x)$ shifts from $k_x = 0$ to $k_x = \pi$ by $2\pi$, which gives rise to a winding number $w=2$.

\bibliography{ref}

\end{document}